\documentclass[preprint,aps]{revtex4}

\usepackage{graphicx}
\newcommand {\bea}{\begin{eqnarray}}
\newcommand {\eea}{\end{eqnarray}}
\newcommand {\be}{\begin{equation}}
\newcommand {\ee}{\end{equation}}

\renewcommand{\a}{\alpha}
\renewcommand{\d}{\delta}
\newcommand{\x}{\times}
\newcommand{\lb}{\lambda}
\newcommand{\g}{\gamma}

\newcommand{\s}{\sigma}
\newcommand{\smn}{\sigma_{\mu\nu}}
\newcommand{\sng}{\sigma_{\nu\gamma}}
\newcommand{\sgm}{\sigma_{\gamma\mu}}

\renewcommand{\r}{\rho}

\newcommand{\noto}{\to\hspace{-0.47cm}/\hspace{0.3cm}}
                      
\begin{document}


\title{Instanton contribution to scalar charmonium and glueball
decays}

\author{V.~Zetocha$^1$ and T.~Sch\"afer$^{1,2}$}

\affiliation{
$^1$Department of Physics, SUNY Stony Brook, Stony Brook, NY 11794\\ 
$^2$Riken-BNL Research Center, Brookhaven National Laboratory, 
Upton, NY 11973}

\begin{abstract}
 We study instanton contributions to hadronic decays of the 
scalar glueball, the pseudoscalar charmonium state $\eta_c$,
and the scalar charmonium state $\chi_c$. Hadronic decays of
the $\eta_c$ are of particular interest. The three main decay
channels are $K\bar{K}\pi$, $\eta\pi\pi$ and $\eta'\pi\pi$, 
each with an unusually large branching ratio $\sim 5\%$. 
On the quark level, all three decays correspond to an instanton
type vertex $(\bar{c}c)(\bar{s}s)(\bar{d}d)(\bar{u}u)$. We show 
that the total decay rate into three pseudoscalar mesons can
be reproduced using an instanton size distribution consistent
with phenomenology and lattice results. Instantons correctly
reproduce the ratio $B(\pi\pi\eta)/B(\pi\pi\eta')$ but 
over-predict the ratio $B(K\bar{K}\pi)/B(\pi\pi\eta(\eta'))$. 
We consider the role of scalar resonances and suggest that the 
decay mechanism can be studied by measuring the angular
distribution of decay products.

\end{abstract}

\maketitle
\newpage

\section{Introduction}

  The charmonium system has played an important role in shaping
our knowledge of perturbative and non-perturbative QCD. The 
discovery of the $J/\psi$ as a narrow resonance in $e^+e^-$ 
annihilation confirmed the existence of a new quantum number,
charm. The analysis of charmonium decays in $e^+e^-$ pairs, 
photons and hadrons established the hypothesis that the 
$J/\psi$ and $\eta_c$ are, to a good approximation, 
non-relativistic $^3S_1$ and $^1S_0$ bound states of heavy 
charm and anti-charm quarks. However, non-perturbative 
dynamics does play an important role in the charmonium
system \cite{Novikov:dq,Shifman:nx}. For example, an analysis 
of the $\psi$ spectrum lead to the first determination of the 
gluon condensate.
  
  The total width of charmonium is dominated by short 
distance physics and can be studied in perturbative 
QCD \cite{Appelquist:zd}. The only non-perturbative input 
in these calculations is the wave function at the origin. 
A systematic framework for these calculations is provided 
by the non-relativistic QCD (NRQCD) factorization method
\cite{Bodwin:1994jh}. NRQCD facilitates higher order calculations 
and relates the decays of states with different quantum numbers. 
QCD factorization can also be applied to transitions of the 
type $\psi'\to\psi+X$ \cite{Gottfried:1977gp,Voloshin:hc}. 

  The study of exclusive decays of charmonium into light 
hadrons is much more complicated and very little work in this 
direction has been done. Perturbative QCD implies some helicity 
selection rules, for example $\eta_c \noto \rho\rho,p\bar{p}$
and $J/\psi \noto \rho\pi,\rho a_1$ 
\cite{Brodsky:1981kj,Chernyak:1983ej}, but these rules 
are strongly violated \cite{Anselmino:yg}. The $J/\psi$ 
decays mostly into an odd number of Goldstone bosons. 
The average multiplicity is $\sim (5-7)$, which is consistent 
with the average multiplicity in $e^+e^-$ annihilation away
from the $J/\psi$ peak. Many decay channels have been 
observed, but none of them stand out. Consequently, we 
would expect the $\eta_c$ to decay mostly into an even 
number of pions with similar multiplicity. However, the 
measured decay rates are not in accordance with this 
expectation. The three main decay channels of the $\eta_c$ 
are $K\bar{K}\pi$, $\eta\pi\pi$ and $\eta'\pi\pi$, each 
with an unusually large branching ratio of $\sim 5$\%. 
Bjorken observed that these three decays correspond to 
a quark vertex of the form $(\bar{c}c)(\bar{s}s)(\bar{d}d)
(\bar{u}u)$ and suggested that $\eta_c$ decays are a 
``smoking gun'' for instanton effects in heavy quark 
decays \cite{Bjorken:2000ni}. 
  
 In this paper we shall try to follow up on this idea by 
performing a more quantitative estimate of the instanton 
contribution to $\eta_c$ and $\chi_c$ decays. The paper
is organized as follows. In section \ref{sec_eff} we 
introduce the instanton induced effective lagrangian. 
In the following sections we apply the effective lagrangian 
to the decays of the scalar glueball, eta charm, and chi 
charm. We should note that this investigation should be 
seen as part of a larger efffort to identify ``direct'' 
instanton contributions in hadronic reactions, such as 
deep inelastic scattering, the $\Delta I=1/2$ rule, or 
$\eta$ production in $pp$ scattering
\cite{Balitsky:1993jd,Moch:1996bs,Kochelev:2001pp,Kochelev:1999tc}.

\section{Effective Lagrangians}
\label{sec_eff}

  Instanton effects in hadronic physics have been studied
extensively \cite{Diakonov:1995ea,Schafer:1996wv}. Instantons 
play an important role in understanding the $U(1)_A$ anomaly 
and the mass of the $\eta'$. In addition to that, there is also 
evidence that instantons provide the mechanism for chiral 
symmetry breaking and play an important role in determining
the structure of light hadrons. All of these phenomena are 
intimately related to the presence of chiral zero modes in
the spectrum of the Dirac operator in the background field 
of an instanton. The situation in heavy quark systems is 
quite different. Fermionic zero modes are not important 
and the instanton contribution to the heavy quark potential
is small \cite{Callan:1978ye}. 

 This does not imply that instanton effects are not relevant. 
The non-perturbative gluon condensate plays an important 
role in the charmonium system \cite{Novikov:dq,Shifman:nx},
and instantons contribute to the gluon condensate. In general, 
the charmonium system provides a laboratory for studying 
non-perturbative glue in QCD. The decay of a charmonium 
state below the $D\bar{D}$ threshold involves an intermediate 
gluonic state. Since the charmonium system is small, $r_{c\bar{c}}
\sim (v m_c)^{-1} < \Lambda_{QCD}^{-1}$, the gluonic system
is also expected to be small. For this reason charmonium decays 
have long been used for glueball searches. 

  Since charmonium decays produce a small gluonic system we
expect that the $c\bar{c}$ system mainly couples to instantons
of size $\rho\sim r_{c\bar{c}}\sim (vm_c)^{-1}$. In this limit 
the instanton effects can be summarized in terms of an effective 
lagrangian \cite{Callan:1977gz,'tHooft:up,Shifman:nz,Shifman:uw}. 
\bea
\label{main_lagr}
{\cal L}_I &=&\int  \prod_q \left[
        m_q\r - 2\pi^2\r^3 \bar{q}_R
	\left(U\mbox{\boldmath{1}}_2U^{\dagger} + 
	\frac{i}{2}t^aR^{aa'}\bar{\eta}^{a'}_{\mu\nu}\s^{\mu\nu}
	\right)q_L  \right] \nonumber \\
 & &\times \exp\left(
	-\frac{2\pi^2}{g}\r^2 \bar{\eta}^{b'}_{\g\delta} R^{b'b}G^{b,\g\d}
 \right)dz\frac{d_0(\r)}{\r^5}d\r\;dU  ,
\eea
where $t^a=\frac{1}{2}\lambda^a$ with ${\rm tr}[\lambda^a\lambda^b]
=2\delta^{ab}$ are $SU(3)$ generators, ${1}_2 = {\rm diag}
(1,1,0)$, $\eta^a_{\mu\nu}$ is the 't Hooft symbol and $\smn=
\frac{1}{2}[\g_\mu,\g_\nu]$. The instanton is characterized by 
$4N_c$ collective coordinates, the instanton position $z$, the 
instanton size $\r$, and the color orientation $U\in SU(N_c)$. 
We also define the rotation matrix $R^{ab}$ by $R^{aa'}\lambda^{a'}
=U\lambda^aU^{\dagger}$. For an anti-instanton we have to replace 
$L\leftrightarrow R$ and $\bar{\eta}\leftrightarrow \eta$. The 
semi-classical instanton density $d(\rho)$ is given by
\be
\label{drho}
  d(\rho) = \frac{d_0(\rho)}{\rho^5}=
   \frac{0.466\exp(-1.679N_c)1.34^{N_f}}{(N_c-1)!(N_c-2)!}\,
  \left(\frac{8\pi^2}{g^2}\right)^{2N_c} 
 \rho^{-5}\exp\left[-\frac{8\pi^2}{g(\rho)^2}\right],
\ee
where $g(\rho)$ is the running coupling constant. For small
$\rho$ we have $d(\rho)\sim \rho^{b-5}$ where $b=(11N_c)/3
-(2N_f)/3$ is the first coefficient of the beta function.

  Expanding the effective lagrangian in powers of the external
gluon field gives the leading instanton contribution to different 
physical matrix elements. If the instanton size is very small,
$\rho\ll m_c^{-1}$, we can treat the charm quark mass as light 
and there is an effective vertex of the form $(\bar{u}u)(\bar{dd})
(\bar{s}s)(\bar{c}c)$ which contributes to charmonium decays. 
Since the density of instantons grows as a large power of $\rho$
the contribution from this regime is very small. In the realistic
case $\rho\sim (vm_c)^{-1}$ we treat the charm quark as heavy 
and the charm contribution to the fermion determinant is 
absorbed in the instanton density $d(\rho)$. The dominant
contribution to charmonium decays then arises from expanding
the gluonic part of the effective lagrangian to second order 
in the field strength tensor. This provides effective vertices
of the form $(G\tilde G)(\bar{u}\gamma_5 u)(\bar{d}\gamma_5 d)
(\bar{s}\gamma_5 s)$, $(G^2)(\bar{u}\gamma_5 u)(\bar{d}\gamma_5
d)(\bar{s}s)$, etc. 

  We observe that the $N_f=3$ fermionic lagrangian combined
with the gluonic term expanded to second order in the field
strength involves an integral over the color orientation of the 
instanton which is of the form $\int dU(U_{ij}U_{kl}^\dagger)^5$.
This integral gives $(5!)^2$ terms. A more manageable result
is obtained by using the vacuum dominance approximation. We
assume that the coupling of the initial charmonium or glueball
state to the instanton proceeds via a matrix element of the 
form $\langle 0^{++}|G^2|0\rangle$ or $\langle 0^{-+}|G\tilde{G}
|0\rangle$. In this case we can use
\be
\langle 0^{++}|G^a_{\mu\nu}G^b_{\alpha\beta}|0\rangle=
 \frac{1}{12(N_c^2-1)}\delta^{ab} 
  (\delta_{\mu\alpha}\delta_{\nu\beta}-\delta_{\mu\beta}
  \delta_{\nu\alpha})
 \langle 0^{++}|G^{a'}_{\rho\sigma}G^{a'}_{\rho\sigma}|0\rangle
\ee 
in order to simplify the color average. The vacuum dominance 
approximation implies that the color average of the fermionic 
and gluonic parts of the interaction can be performed independently. 
In the limit of massless quarks the instanton ($I$) and anti-instanton
($A$) lagrangian responsible for the decay of scalar and pseudoscalar 
charmonium decays is given by
\be
\label{main}
{\cal L}_{I+A} = \int\!\!dz\frac{d_0(\r)}{\r^5}d\r
\frac{\pi^3\r^4}{(N_c^2-1)\alpha_s} 
\left\{
 \left(G^2-G\tilde{G}\right)\times L_{f,I} +
 \left(G^2+G\tilde{G}\right)\times L_{f,A}
\right\}. 
\ee
Here, ${\cal L}_{f,IA}$ is the color averaged $N_f=3$ fermionic 
effective lagrangian \cite{Shifman:uw,Diakonov:1995ea,Schafer:1996wv}.

\section{Scalar glueball decays}
\label{sec_glue}

  Since the coupling of the charmonium state to the instanton
proceeds via an intermediate gluonic system with the quantum 
numbers of scalar and pseudoscalar glueballs it is natural
to first consider direct instanton contributions
to glueball decays. This problem is of course important in its 
own right. Experimental glueball searches have to rely on identifying
glueballs from their decay products. The successful identification
of a glueball requires theoretical calculations of glueball mixing
and decay properties. In the following we compute the direct
instanton contribution to the decay of the scalar $0^{++}$ 
glueball state into $\pi\pi$, $K\bar{K}$, $\eta\eta$ and 
$\eta\eta'$.

 Since the initial state is parity even only the $G^2$ term in
equ.~(\ref{main}) contributes. The relevant effective interaction
is given by
\bea
\label{glue_lagr}
  {\cal L}_{I+A}
   &=&\int\!\! dz\int\!\! 
	d_0(\rho)\frac{d\rho}{\rho^5}\frac{1}{N_c^2-1} 
	\left( 
	  \frac{\pi^3\rho^4}{\alpha_s}\right) G^2 
	  (-\frac{1}{4}) \left( \frac{4}{3}\pi^2\rho^3 
	\right)^3 \times  \nonumber \\
 &  &\biggl\{
	[
	(\bar{u}u)(\bar{d}d)(\bar{s}s) +
	(\bar{u}\gamma^5u)(\bar{d}\gamma^5d)(\bar{s}s) +
	(\bar{u}\gamma^5u)(\bar{d}d)(\bar{s}\gamma^5s) +
	(\bar{u}u)(\bar{d}\gamma^5d)(\bar{s}\gamma^5s) 
	]     \nonumber \\
 &  &+\frac{3}{8}
      {\biggl[}
	(\bar{u}t^au)(\bar{d}t^ad)(\bar{s}s) +
	(\bar{u}t^a \gamma^5u)(\bar{d}t^a \gamma^5d)(\bar{s}s) +
	(\bar{u}t^a \gamma^5u)(\bar{d}t^a d)(\bar{s}\gamma^5s)  
           \nonumber \\
 &  & \hspace{0.5cm}
       +(\bar{u}t^a u)(\bar{d}t^a\gamma^5d)(\bar{s}\gamma^5s) \nonumber\\
 & &  \hspace{0.1cm} 
       -\frac{3}{4}\bigl[
	(\bar{u}t^a \sigma_{\mu\nu} u)(\bar{d}t^a \smn d)(\bar{s}s) +
	(\bar{u}t^a \smn \gamma^5u)(\bar{d}t^a \smn\gamma^5d)(\bar{s}s)
           \nonumber \\
 &  &	\hspace{0.5cm}
      +(\bar{u}t^a \smn\gamma^5u)(\bar{d}t^a\smn d)(\bar{s}\gamma^5s) +
	(\bar{u}t^a\smn u)(\bar{d}t^a\smn\gamma^5d)(\bar{s}\gamma^5s)\bigr]  
            \nonumber \\
 &  & \hspace{0.1cm}
       -\frac{9}{20}d^{abc} \bigl[ 
	(\bar{u}t^a \sigma_{\mu\nu} u)(\bar{d}t^b \smn d)(\bar{s}t^cs) +
	(\bar{u}t^a \smn \gamma^5u)(\bar{d}t^b\smn\gamma^5d)(\bar{s}t^cs) 
            \nonumber \\
 &  & \hspace{0.5cm}
       +(\bar{u}t^a \smn\gamma^5u)(\bar{d}t^b\smn d)(\bar{s}t^c\gamma^5s) +
	(\bar{u}t^a\smn u)(\bar{d}t^b \smn\gamma^5d)(\bar{s}t^c\gamma^5s)  
       \bigr]\nonumber \\
 &  &  \hspace{0.5cm} + (2\; {\it cyclic\; permutations}\; 
           u\leftrightarrow d \leftrightarrow s)
      {\biggr]}\nonumber \\
 &  & \hspace{0.1cm}-\frac{9}{40}d^{abc}
       \biggl[  
	(\bar{u}t^au)(\bar{d}t^bd)(\bar{s}t^cs) +
	(\bar{u}t^a\gamma^5u)(\bar{d}t^b\gamma^5d)(\bar{s}t^cs) +
	(\bar{u}t^a\gamma^5u)(\bar{d}t^bd)(\bar{s}\gamma^5t^cs)   
                   \nonumber \\
 &  &	\hspace{0.5cm}
       +(\bar{u}t^au)(\bar{d}t^b\gamma^5d)(\bar{s}t^c\gamma^5s) 
         \biggr] \nonumber  \\
 &  & \hspace{0.1cm} 
       - \frac{9}{32}if^{abc} \biggl[   
	(\bar{u}t^a \sigma_{\mu\nu} u)(\bar{d}t^b \sng d)
	(\bar{s}t^c\sgm s)  
       +(\bar{u}t^a \smn\gamma^5u)(\bar{d}t^b\sng\gamma^5d) 
	(\bar{s}t^c\sgm s) \nonumber \\
 &  & \hspace{0.5cm}
       + (\bar{u}t^a\smn\gamma^5u)(\bar{d}t^b\sng d) 
	(\bar{s}t^c\sgm\gamma^5s) 
       +(\bar{u}t^a\smn u)
	(\bar{d}t^b\sng\gamma^5d)(\bar{s}t^c\sgm\gamma^5s)  
      \biggr]
   \biggr\}  
\eea
Let us start with the process $0^{++} \rightarrow \pi\pi$. 
In practice we have Fierz rearranged equ.~(\ref{glue_lagr}) 
into structures that involve the strange quark condensate 
$\bar{s}s$ as well as operators with the quantum numbers of 
two pions. In order to compute the coupling of these operators 
to the pions in the final state we have used PCAC relations
\bea
\langle 0|\bar{d}\g^5u|\pi^+\rangle
  &=& \frac{i\sqrt{2}m_\pi^2f_\pi}{m_u+m_d}\equiv K_\pi ,  \\
\langle0|\bar{s}\g^5u|K^+\rangle
  &=&\frac{i\sqrt{2}m_{K}^2f_K}{m_u+m_s}\equiv K_{K}  .
\eea
The values of the decay constants are $f_\pi=93$ MeV, $f_K=113$ 
MeV \cite{PDG}. We also use $Q_u\equiv \langle\bar{u}u\rangle = 
-(248\,{\rm MeV})^3$ and $Q_d=Q_u$ as well as $Q_s=0.66Q_u$ 
\cite{Narison:2002hk}. The coupling of the $\eta'$ meson is 
not governed by chiral symmetry. A recent analysis of $\eta-\eta'$ 
mixing and the chiral anomaly gives \cite{Feldmann:1999uf} 
\bea
 \langle 0|\bar{u}\gamma_5 u |\eta\rangle  &=& 
    \langle 0|\bar{d}\gamma_5 d|\eta\rangle = 
      - i(358\; {\rm  MeV} )^2\equiv K_\eta^q , \\
 \langle 0|\bar{u}\gamma_5 u |\eta'\rangle  &=& 
     \langle 0|\bar{d}\gamma_5 d|\eta'\rangle = 
      - i(320\; {\rm  MeV} )^2\equiv K_{\eta'}^q ,\nonumber\\
 \langle 0|\bar{s}\gamma_5 s| \eta\rangle &=& i(435\;{\rm MeV})^2 
       \equiv K_{\eta}^s ,\nonumber \\
 \langle 0|\bar{s}\gamma_5 s| \eta'\rangle &=& 
        -i(481\;{\rm MeV})^2 \equiv K_{\eta'}^s . \nonumber
\eea
Finally, we need the coupling of the glueball state to the 
gluonic current. This quantity has been estimated using QCD 
spectral sum rules \cite{Novikov:va,Narison:1996fm}
and the instanton model \cite{Schafer:1994fd}. We use
\be
 \langle0^{++}|g^2G^2|0\rangle
\equiv \lb_0 = 15\,{\rm GeV}^3.
\ee
We can now compute the matrix element for $0^{++}\rightarrow
\pi^+\pi^-$. The interaction vertex is 
\be
\label{0++pipi}
{\cal L}_{I+A}^{\pi^+\pi^-}
  = \int\!\! dz
   \int\!\!\frac{d\r}{\r^5}d_0(\r)
  \frac{1}{N_c^2-1}\left(\frac{\pi^3\r^4}{\alpha_s^2}\right)
\left(
\frac{4}{3}\pi^2\r^3
\right)^3 
 \x \frac{1}{4}
   (\a_sG^2)(\bar{s}s)(\bar{u}\g^5d)(\bar{d}\g^5u). 
\ee
The integral over the position of the instanton leads to a momentum 
conserving delta function, while the vacuum dominance approximation
allows us to write the amplitude in terms of the coupling constants
introduced above. We find
\be
\langle0^{++}(q)|\pi^+(p^+)\pi^-(p^-)\rangle
    = (2\pi)^4\delta^4(q-p^+-p^-)
      \frac{A}{16\pi}\lb_0 Q_s K_\pi^2,
\ee
where 
\be
\label{A_int}
 A=\int\frac{d\r}{\r^5}d_0(\r)\frac{1}{N_c^2-1}
 \left(\frac{\pi^3\r^4}{\a_s^2}\right) 
 \left(\frac{4}{3}\pi^2\r^3\right)^3.
\ee
The instanton density $d_0(\rho)$ is known accurately only
in the limit of small $\r$. For large $\rho$ higher loop 
corrections and non-perturbative effects are important.
The only source of information in this regime is lattice QCD 
\cite{Michael:1995br,Smith:1998wt,deForcrand:1997sq,DeGrand:1997gu}.
A very rough caricature of the lattice results is provided
by the parameterization
\be
\frac{d_0(\r)}{\r^5}=\frac{1}{2}n_0\d(\r-\r_c),
\ee
with $n_0 \simeq 1\,{\rm fm}^{-4}$ and $\r_c \simeq 0.33\,
{\rm fm}$. This parameterization gives $A=(379\,{\rm MeV})^{-9}$.
Another way to compute $A$ is to regularize the integral over
the instanton size by replacing $d(\rho)$ with $d(\rho)\exp(-
\alpha\rho^2)$. The parameter $\alpha$ can be adjusted in order 
to reproduce the size distribution measured on the lattice. 
We notice, however, that whereas the instanton density scales 
as $\rho^{b-5}\sim \rho^4$, the decay amplitude scales as 
$\rho^{b+8}\sim \rho^{17}$. This implies that the results 
are very sensitive to the density of large instantons. We 
note that when we study the decay of a small-size bound 
state the integral over $\rho$ should be regularized by 
the overlap with the bound state wave function. We will 
come back to this problem in section \ref{sec_eta} below. 

\begin{figure}
\begin{center}
\includegraphics[width=10cm,angle=-90]{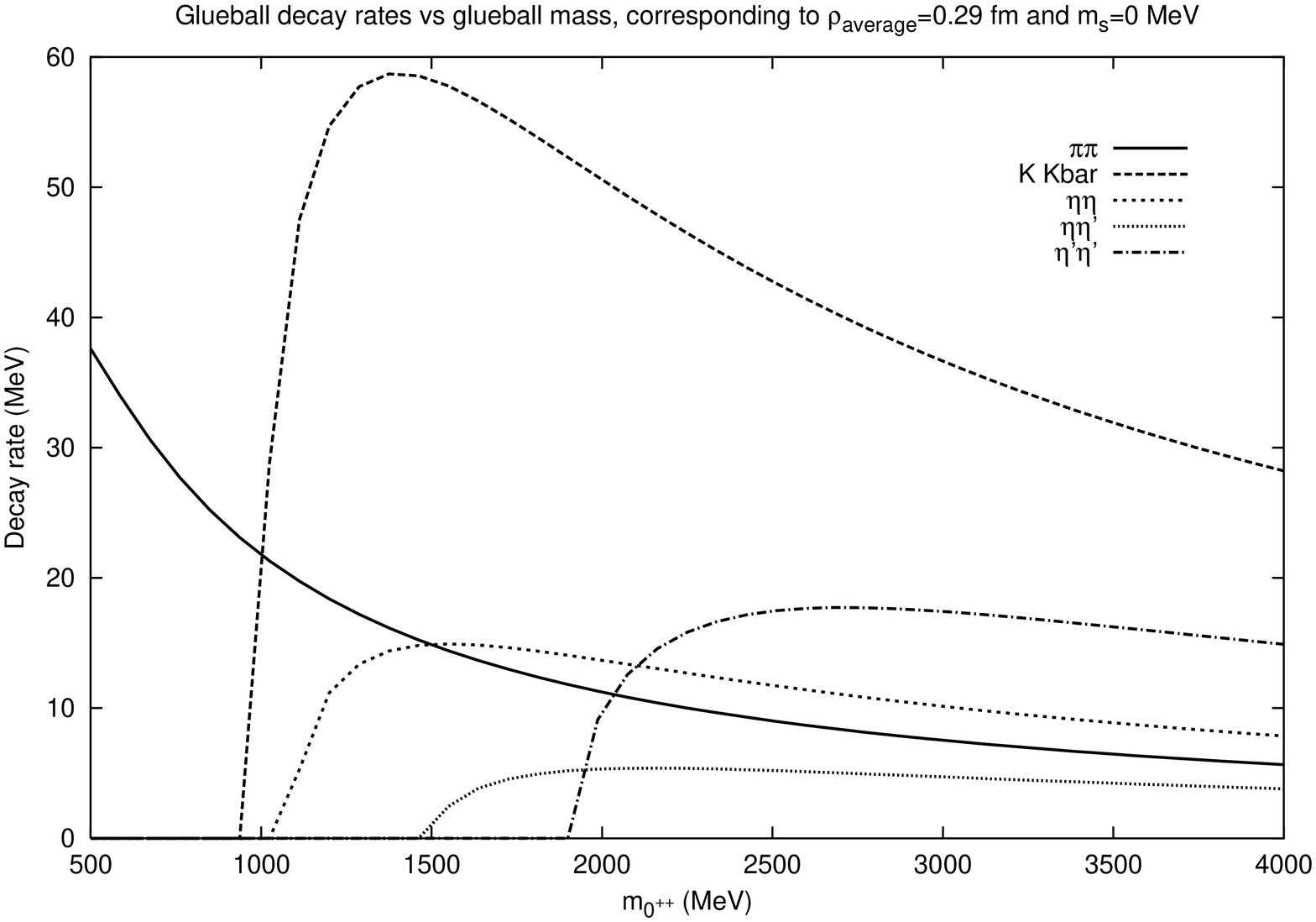}
\end{center}
\begin{center}
\caption{\label{fig_0++}
Scalar glueball decay rates plotted a function of the mass 
of the scalar glueball. The rates shown in this figure were
computed from the instanton vertex in the chiral limit. The 
average instanton size was taken to be $\bar{\rho}=0.29$ fm.}
\end{center}
\end{figure}

 We begin by studying ratios of decay rates. These ratios 
are not sensitive to the instanton size distribution. The 
decay rate $0^{++}\to\pi^+\pi^-$ is given by
\be
\Gamma_{0^{++}\rightarrow \pi^+\pi^-}
 =\frac{1}{16\pi}\frac{\sqrt{m_{0^{++}}^2-4m_{\pi}^2}}{m_{0^{++}}^2}
 \left[ \frac{A}{16\pi}\lb_0Q_sK_{\pi}^2 \right]^2 .
\ee
The decay amplitude for the process $0^{++}\rightarrow\pi_0\pi_0$ 
is equal to the $0^{++}\rightarrow\pi^+\pi^-$ amplitude as required 
by isospin symmetry. Taking into account the indistinguishability 
of the two $\pi_0$ we get the total $\pi\pi$ width
\be
\Gamma_{0^{++}\rightarrow \pi\pi} =
  \frac{3}{32\pi}
   \frac{\sqrt{m_{0^{++}}^2-4m_{\pi}^2}}{m_{0^{++}}^2}
   \left[\frac{A}{16\pi}\lb_0Q_sK_{\pi}^2\right]^2 .
\ee
In a similar fashion we get the decay widths for the $K\bar{K}$, 
$\eta\eta$, $\eta\eta'$ and $\eta'\eta'$ channels
\bea
\Gamma_{0^{++}\rightarrow K\bar{K}}
  &=& 2\frac{1}{16\pi}
   \frac{\sqrt{m_{0^{++}}^2-4m_{K}^2}}{m_{0^{++}}^2}
    \left[ \frac{A}{16\pi}\lb_0Q_uK_K^2 \right]^2,  \\
\Gamma_{0^{++}\rightarrow \eta\eta}
  &=&  \frac{1}{32\pi}
  \frac{\sqrt{m_{0^{++}}^2-4m_{\eta}^2}}{m_{0^{++}}^2}
  \left[ \frac{A}{16\pi}\lb_0
    K_{\eta}^q\;2(Q_sK_{\eta}^q+(Q_u+Q_d)K_{\eta}^s)
  \right]^2,   \\
\Gamma_{0^{++}\rightarrow \eta\eta'}
 &=&\frac{1}{16\pi}
  \frac{\sqrt{[m_{0^{++}}^2-(m_{\eta}+m_{\eta'})^2]
  [m_{0^{++}}^2-(m_\eta-m_{\eta'})^2]}}{m_{0^{++}}^3} \nonumber \\
  & & \x \left[
   \frac{A}{16\pi}\lb_0(2Q_sK_{\eta}^qK_{\eta'}^q+
   (Q_u+Q_d)(K_{\eta}^qK_{\eta'}^s+K_{\eta}^sK_{\eta'}^q)
   \right]^2  \\
\Gamma_{0^{++}\rightarrow \eta'\eta'}
 &=&  \frac{1}{32\pi}
   \frac{\sqrt{m_{0^{++}}^2-4m_{\eta'}^2}}{m_{0^{++}}^2}
   \left[ \frac{A}{16\pi}\lb_0 K_{\eta'}^q2(Q_sK_{\eta'}^q
       +(Q_u+Q_d)K_{\eta'}^s) \right]^2 .
\eea
Here, $\bar{K}K$ refers to the sum of the $K^+K^-$ and 
$\bar{K}_0K_0$ final states. We note that in the chiral 
limit the instanton vertices responsible for $\pi\pi$ 
and $\bar{K}K$ decays are identical up to quark interchange. 
As a consequence, the ratio of the decay rates $\Gamma_{0^{++}
\rightarrow \pi\pi}/ \Gamma_{0^{++}\rightarrow K\bar{K}}$ 
is given by the phase space factor multiplied by the ratio 
of the coupling constants
\be
\label{0++ratio}
\frac{\Gamma_{0^{++}\rightarrow \pi\pi}} 
     {\Gamma_{0^{++}\rightarrow K\bar{K}}}
   = \frac{3}{4}\x \frac{Q_s^2 K_\pi^4}{Q_u^2 K_K^4}\x
      \sqrt{\frac{m_{0^{++}}^2-4m_{\pi}^2}{m_{0^{++}}^2-4m_K^2}}=
  (0.193 \pm 0.115)
      \sqrt{\frac{m_{0^{++}}^2-4m_{\pi}^2}{m_{0^{++}}^2-4m_K^2}}.
\ee
The main uncertainty in this estimate comes from the value of 
$m_s$, which is not very accurately known. We have used 
$m_s=(140\pm 20)\,{\rm MeV}$. The ratio of $\pi\pi$ to $\eta\eta$ 
decay rates is not affected by this uncertainty,
\be
\label{0++ratio_pi_eta}
\frac{\Gamma_{0^{++}\rightarrow \pi\pi}} 
     {\Gamma_{0^{++}\rightarrow \eta\eta}}
  =  0.69
   \sqrt{\frac{m_{0^{++}}^2-4m_{\pi}^2}{m_{0^{++}}^2-4m_\eta^2}}.
\ee
In Fig.\ref{fig_0++} we show the decay rates as functions of 
the glueball mass. We have used $\Lambda_{QCD}=300\,{\rm MeV}$ 
and adjusted the parameter $\alpha$ to give the average instanton 
size $\bar{\rho}=0.29$ fm. We observe that for glueball masses
$m_{0^{++}}>1$ GeV the $K\bar{K}$ phase space suppression 
quickly disappears and the total decay rate is dominated
by the $K\bar{K}$ final state. We also note that for 
$m_{0^{++}}>1.5$ GeV the $\eta\eta$ rate dominates over
the $\pi\pi$ rate. 

\begin{figure}
\begin{center}
\includegraphics[width=10cm,angle=-90]{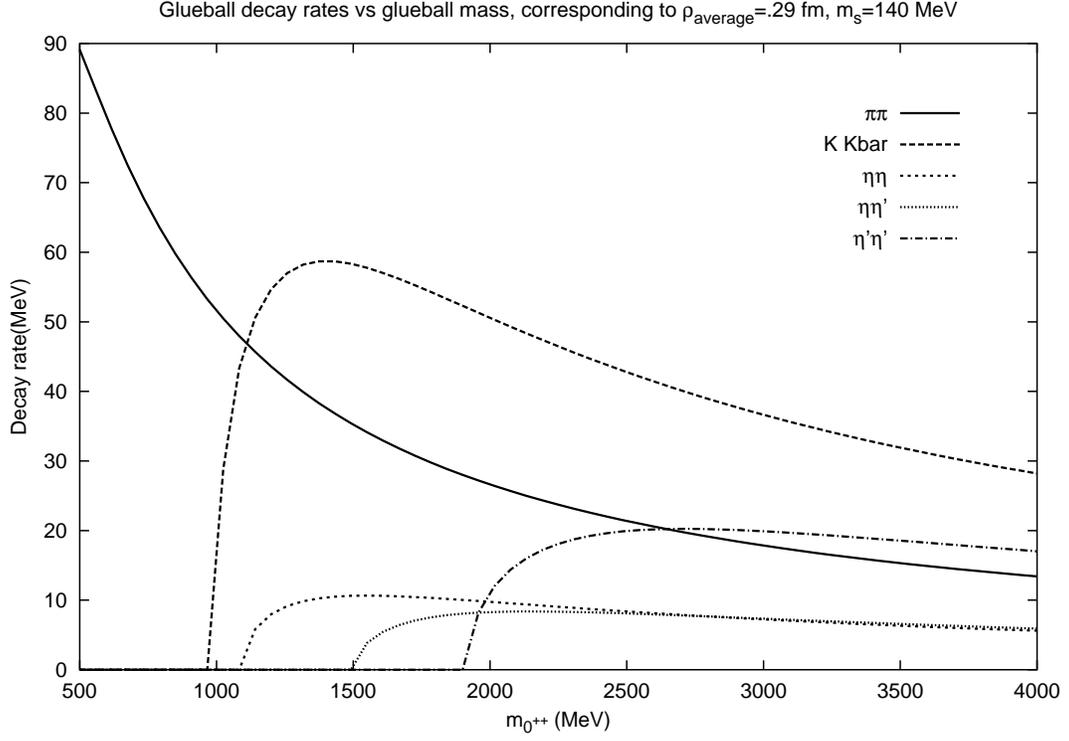}
\end{center}
\begin{center}
\caption{\label{fig_0++_ms}
Same as Fig.~\ref{fig_0++} but with $m_s\neq 0$ corrections
in the instanton vertex taken into account. The results 
shown in this figure correspond to $m_s=140$ MeV.}
\end{center}
\end{figure}

\begin{figure}
\begin{center}
\includegraphics[width=10cm,angle=-90]{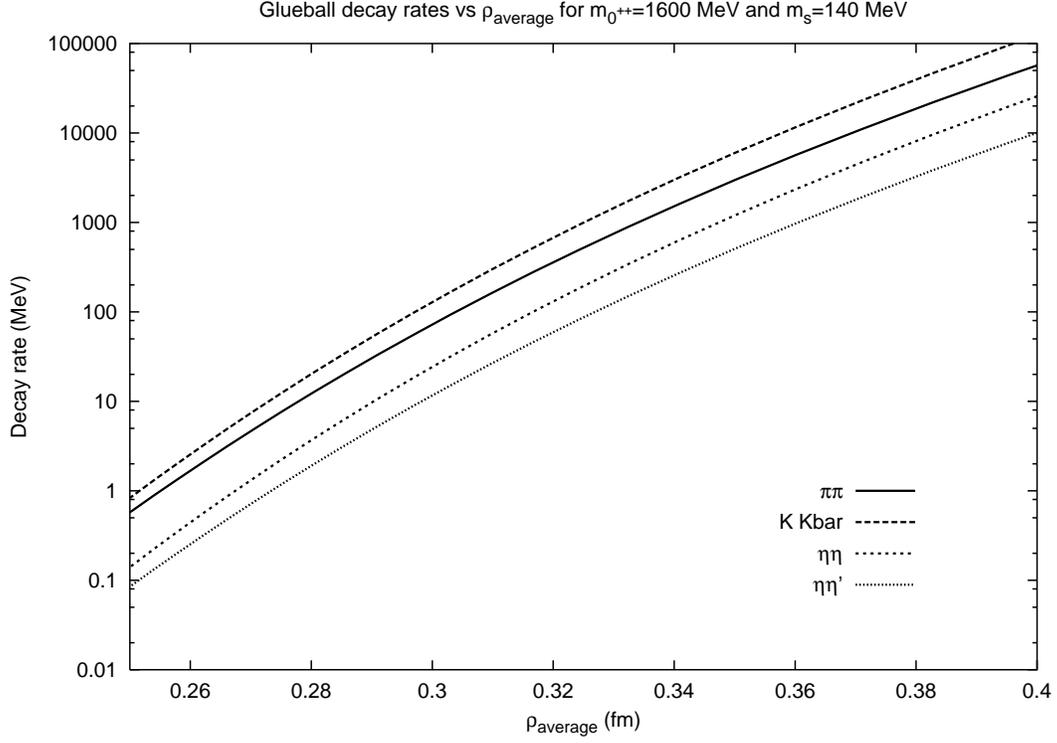}
\end{center}
\begin{center}
\caption{\label{fig_glue_rho_ms}
Dependence of glueball decay rates on the average instanton size.
The results shown in this figure correspond to the instanton
vertex with $m_s\neq 0$ terms included. The strange quark mass
was taken to be $m_s=140$ MeV.}
\end{center}
\end{figure}

 In deriving the effective instanton vertex equ.~(\ref{0++pipi}) 
we have taken all quarks to be massless. While this is a good 
approximation for the up and down quarks, this it is not necessarily 
the case for the strange quark. The $m_s\ne 0$ contribution to the 
effective interaction for $0^{++}$ decay is given by
\bea
 {\cal L}_{m_s} 
  &=& \int\, \frac{d\rho}{\rho^5}d_0(\rho) \frac{1}{N_c^2-1}
     \frac{\pi^3\rho^4}{\alpha_s^2}\left(\frac{4}{3}\pi^2\rho^3 \right)^2
      m_s\rho (\alpha_sG^2)\x  \\
  & & \frac{1}{2} 
   \left\{ 
     (\bar{u}u)(\bar{d}d) + (\bar{u}\gamma^5u)(\bar{d}\gamma^5d) +
   \frac{3}{8} \left[
     (\bar{u}t^au)(\bar{d}t^ad) + 
     (\bar{u}\gamma^5t^au)(\bar{d}\gamma^5t^ad)\right. + 
    \right.    \nonumber \\
  & & \mbox{}\hspace{0.2cm}-
    \left.\left.
     \frac{3}{4}(\bar{u}\sigma_{\mu\nu}t^au)(\bar{d}\sigma_{\mu\nu}t^ad) 
    - \frac{3}{4}(\bar{u}\sigma_{\mu\nu}\gamma^5t^au)  
       (\bar{d}\sigma_{\mu\nu}\gamma^5t^a d) 
      \right]  \right\}. \nonumber 
\eea
There is no $m_s\neq0$ contribution to the $K\bar{K}$ channel.
The $m_s\neq 0$ correction to the other decay channels is
\bea
\label{glue_pipi_ms} 
\Gamma_{0^{++}\rightarrow \pi\pi} 
  &=& \frac{3}{32\pi}
   \frac{\sqrt{m_{0^{++}}^2-4m_{\pi}^2}}{m_{0^{++}}^2}
   \left[\frac{1}{16\pi}\lb_0K_{\pi}^2(AQ_s-2Bm_s)\right]^2, \\
\Gamma_{0^{++}\rightarrow \eta\eta}
  &=&\frac{1}{32\pi}
  \frac{\sqrt{m_{0^{++}}^2-4m_{\eta}^2}}{m_{0^{++}}^2}
  \left[\frac{1}{16\pi}\lb_0\;2
  [(AQ_s-2Bm_s)(K_{\eta}^q)^2+A(Q_u+Q_d)K_{\eta}^sK_{\eta}^q)]
  \right]^2 , \nonumber \\
\Gamma_{0^{++}\rightarrow \eta\eta'}
 &=&\frac{1}{16\pi}\frac{\sqrt{[m_{0^{++}}^2-(m_{\eta}+m_{\eta'})^2]
        [m_{0^{++}}^2-(m_\eta-m_{\eta'})^2]}}{m_{0^{++}}^3} \nonumber \\
 & & \x \left[
     \frac{1}{16\pi}\lb_0[2\;(AQ_s-2Bm_s)K_{\eta}^qK_{\eta'}^q
     +A(Q_u+Q_d)(K_{\eta}^qK_{\eta'}^s+K_{\eta'}^qK_{\eta}^s)]
     \right]^2 ,  \nonumber \\
\label{glue_etaPetaP_ms}
\Gamma_{0^{++}\rightarrow \eta'\eta'}
 &=& \frac{1}{32\pi}
   \frac{\sqrt{m_{0^{++}}^2-4m_{\eta'}^2}}{m_{0^{++}}^2}
    \left[
    \frac{1}{16\pi}\lb_02[(AQ_s-2Bm_s)(K_{\eta'}^q)^2
    +A(Q_u+Q_d)K_{\eta'}^sK_{\eta'}^q]
    \right]^2, \nonumber
\eea
where 
\be
\label{B_int}
B=\int\frac{d\r}{\r^5}d_0(\r)\frac{1}{N_c^2-1}
  \left(\frac{\pi^3\r^4}{\a_s^2}\right) 
  \left(\frac{4}{3}\pi^2\r^3\right)^2 \r \; .
\ee
The decay rates with the $m_s\neq 0$ correction to the instanton
vertex taken into account are plotted in Fig.~\ref{fig_0++_ms}.
We observe that effects due to the finite strange quark are 
not negligible. We find that the $\pi\pi$ , $\eta\eta'$, and $\eta'\eta'$
channels are enhanced 
whereas the $\eta\eta$ channel
is reduced. For a typical glueball mass $m_{0^{++}}=(1.5-1.7)$
GeV the ratio $r=B(\pi\pi)/B(K\bar{K})$ changes from $r\simeq 
0.25$ in the case $m_s=0$ to $r\simeq 0.55$ for $m_s\neq 0$. In 
Fig.~\ref{fig_glue_rho_ms} we show the dependence of the decay 
rates on the average instanton size $\bar{\rho}$. We observe 
that using the phenomenological value $\bar{\rho}=0.3$ fm gives 
a total width $\Gamma_{0^{++}}\simeq 100$ MeV. We note, however, 
that the decay rates are very sensitive to the value of $\bar{\rho}$. 
As a consequence, we cannot reliably predict the total decay rate. 
On the other hand, the ratio of the decay widths for different 
final states does not depend on $\bar{\rho}$ and provides a 
sensitive test for the importance of direct instanton effects.

\begin{table}
\begin{tabular}{|c|c|c|c|}\hline\hline
  resonance &   full width   $\Gamma({\rm MeV})$   
  & Mass (MeV) & decay channels \\ 
\hline\hline
$f_0(1370)$ & 200-500 & 1200-1500 & 
  \begin{tabular}{c}
	$\r\r$ dominant \\
	$\pi\pi,K\bar{K},\eta\eta$ seen \\ 
  \end{tabular}
\\ \hline
$f_0(1500)$ & $109\pm 7$ & $1507\pm5$ & 
  \begin{tabular}{lcl}
     $\frac{\Gamma_{K\bar{K}} }{\Gamma_{\pi\pi}}$&=& $0.19 \pm 0.07 $ \\
     $\frac{\Gamma_{\eta\eta'}}{\Gamma_{\pi\pi}}$&=& $0.095 \pm 0.026 $\\ 
     $\frac{\Gamma_{\eta\eta}}{\Gamma_{\pi\pi}}$ &=& $0.18 \pm 0.03  $\\
  \end{tabular}
  \\ \hline
$f_0(1710)$ & $125 \pm 10$ & $ 1713\pm 6$ & 
  \begin{tabular}{lcl}
     $\frac{\Gamma{\pi\pi}}{\Gamma_{K\bar{K}} }$  &=& $0.39 \pm 0.14 $ \\
     $\frac{\Gamma_{\eta\eta}}{\Gamma_{K\bar{K}}}$&=& $0.48 \pm 0.15 $\\ 
  \end{tabular}
\\ \hline
\end{tabular}
\caption{
\label{table_glue}
Masses, decay widths, and decay channels for 
scalar-isoscalar mesons with masses in the $(1-2)$ GeV
range. The data were taken from \cite{PDG}.}
\end{table}

  In Tab.~\ref{table_glue} we show the masses and decay widths 
of scalar-isoscalar mesons in the (1-2) GeV mass range. These
states are presumably mixtures of mesons and glueballs. This 
means that our results cannot be directly compared to experiment
without taking into account mixing effects. It will be interesting
to study this problem in the context of the instanton model, but 
such a study is beyond the scope of this paper. It is nevertheless
intriguing that the $f_0(1710)$ decays mostly into $K\bar{K}$. 
Indeed, a number of authors have suggested that the $f_0(1710)$ 
has a large glueball admixture 
\cite{Sexton:1995kd,Lee:1999kv,Minkowski:1998mf,Close:1996yc}.

\section{Eta charm decays}
\label{sec_eta}

  The $\eta_c$ is a pseudoscalar $J^{PC}=0^{-+}$ charmonium
bound state with a mass $m_{\eta_c}=(2979\pm 1.8)$ MeV. The total 
decay width of the $\eta_c$ is $\Gamma_{\eta_c}=(16\pm 3)$ MeV.
In perturbation theory the total width is given by
\be
 \Gamma(\eta\to 2g) = \frac{8\pi\alpha^2_s|\psi(0)|^2}{3m_c^2} 
   \left( 1 + 4.4\frac{\alpha_s}{\pi}\right).
\ee
Here, $\psi(0)$ is the $^1S_0$ ground state wave function at the
origin. Using $m_c=1.25$ GeV and $\alpha_s(m_c)=0.25$ we get 
$|\psi(0)|\simeq 0.19\,{\rm GeV}^{3/2}$, which is consistent with 
the expectation from phenomenological potential models. Exclusive
decays cannot be reliably computed in perturbative QCD. As
discussed in the introduction Bjorken pointed out that $\eta_c$
decays into three pseudoscalar Goldstone bosons suggest that 
instanton effects are important \cite{Bjorken:2000ni}. The 
relevant decay channels and branching ratios are $B(K\bar{K}\pi)
=(5.5\pm 1.7)\%$, $B(\eta\pi\pi)=(4.9\pm1.8)\%$ and $B(\eta'\pi
\pi)=(4.1\pm 1.7\%)$. These three branching ratios are anomalously 
large for a single exclusive channel, especially given the 
small multiplicity. The total decay rate into these three 
channels is $(14.5\pm 5.2)\%$ which is still a small fraction 
of the total width. This implies that the assumption that 
the three-Goldstone bosons channels are instanton dominated 
is consistent with our expectation that the total width is 
given by perturbation theory. For comparison, the next most 
important decay channels are $B(2(\pi^+\pi^-))=(1.2\pm 0.4)\%$ 
and $B(\r\r)=(2.6\pm 0.9)\%$. These channels do not receive 
direct instanton contributions. 

\begin{figure}
\begin{center}
\includegraphics[width=9cm]{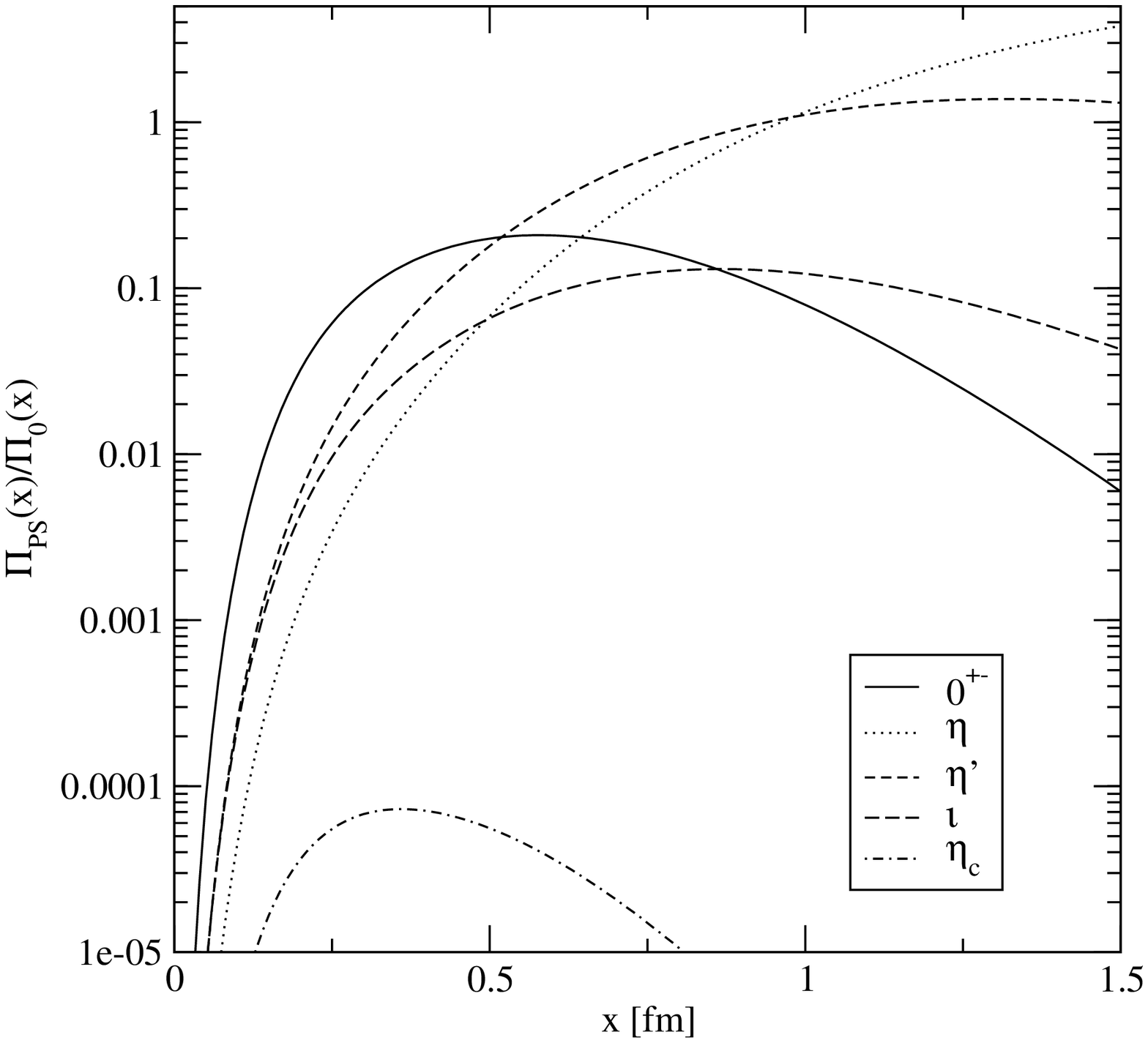}
\includegraphics[width=9cm]{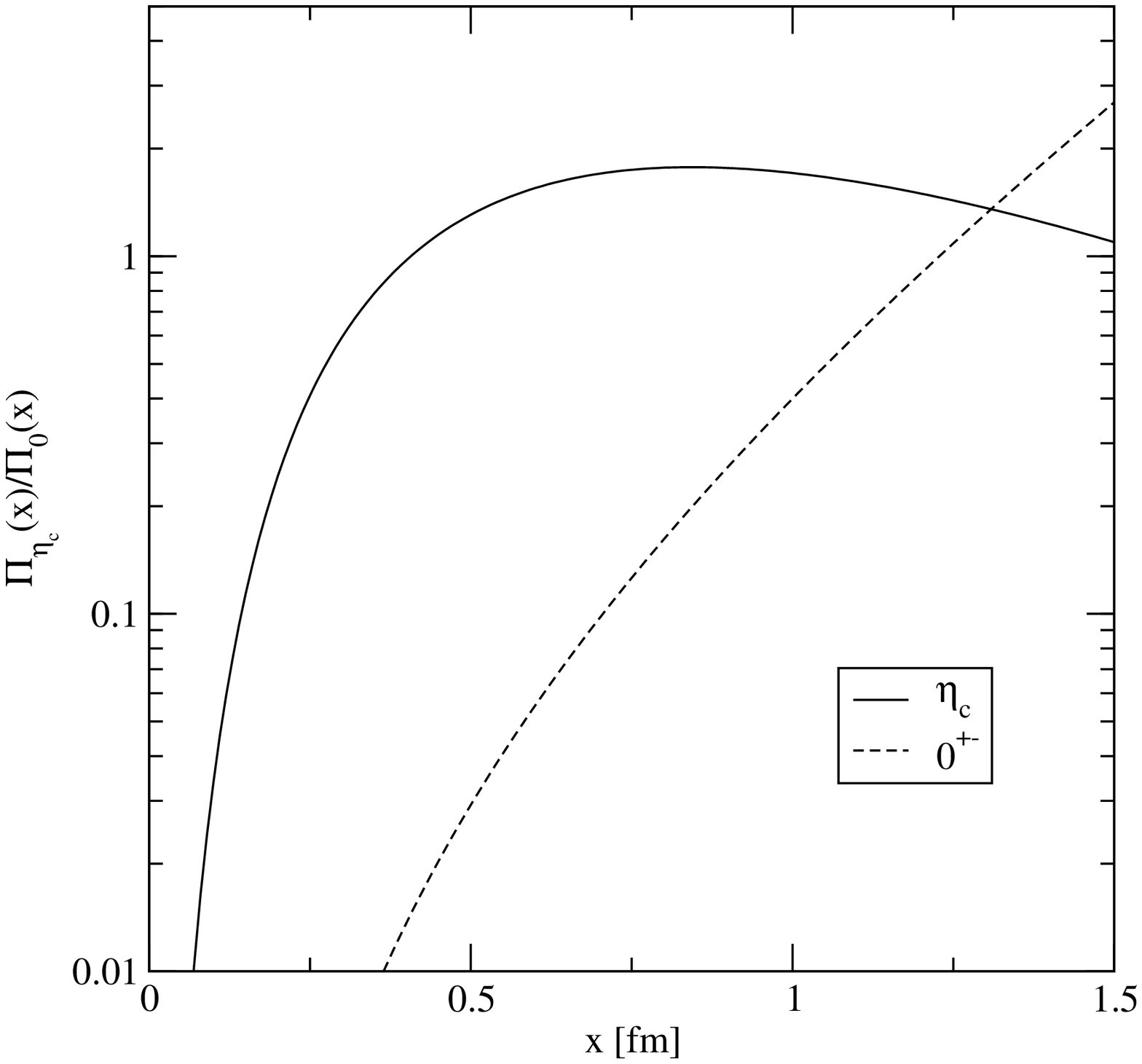}
\end{center}
\begin{center}
\caption{\label{fig_gg}
Resonance contributions to the pseudoscalar glueball 
correlation function $\langle g^2G\tilde{G}(0)g^2G\tilde{G}
(x)\rangle$ and the charmonium correlator $\langle \bar{c}
\gamma_5 c(0)\bar{c}\gamma_5 c(x)\rangle$. Both correlation
functions are normalized to free field behavior. In the 
case of the gluonic correlation function we show the 
glueball contribution compared to the $\eta$, $\eta'$,
$\eta(1440)$ and $\eta_c$ contribution. For the charmonium
correlation function we show the $\eta_c$ and glueball contribution.}
\end{center}
\end{figure}

  The calculation proceeds along the same lines as the glueball
decay calculation. Since the $\eta_c$ is a pseudoscalar only the 
$G\tilde{G}$ term in equ.~(\ref{main}) contributes. The relevant 
interaction is 
\bea
\label{eta_c_lagr}
 {\cal L}_{I+A}
   &=& \int\!\! dz\int\!\! 
	d_0(\rho)\frac{d\rho}{\rho^5}\frac{1}{N_c^2-1} 
	\left( 
	  \frac{\pi^3\rho^4}{\alpha_S}\right) G\tilde{G} 
	  (\frac{1}{4}) \left( \frac{4}{3}\pi^2\rho^3 
	\right)^3 \times  \nonumber \\
 &  &\biggl\{
	[
	(\bar{u}\gamma^5u)(\bar{d}d)(\bar{s}s) +
	(\bar{u}u)(\bar{d}\gamma^5d)(\bar{s}s) +
	(\bar{u}u)(\bar{d}d)(\bar{s}\gamma^5s) +
	(\bar{u}\gamma^5u)(\bar{d}\gamma^5d)(\bar{s}\gamma^5s) 
	]     \nonumber \\
 &  &+\frac{3}{8}
      {\biggl[}
	(\bar{u}t^a\gamma^5u)(\bar{d}t^ad)(\bar{s}s) +
	(\bar{u}t^a u)(\bar{d}t^a \gamma^5d)(\bar{s}s) +
	(\bar{u}t^a u)(\bar{d}t^a d)(\bar{s}\gamma^5s)   \nonumber \\
 &  &  \hspace{0.5cm}
       +(\bar{u}t^a\gamma^5 u)(\bar{d}t^a\gamma^5d)(\bar{s}\gamma^5s) 
        \nonumber \\
 &  &  - \frac{3}{4}\bigl[
	(\bar{u}t^a \sigma_{\mu\nu}\gamma^5 u)(\bar{d}t^a \smn d)(\bar{s}s) +
	(\bar{u}t^a \smn u)(\bar{d}t^a \smn\gamma^5d)(\bar{s}s) 
          \nonumber \\ 
 &  &  \hspace{0.5cm}+
       (\bar{u}t^a \smn u)(\bar{d}t^a\smn d)(\bar{s}\gamma^5s)
	+ (\bar{u}t^a\smn\gamma^5u)
	(\bar{d}t^a\smn\gamma^5d)(\bar{s}\gamma^5s)\bigr]  \nonumber \\
 &  &	-\frac{9}{20}d^{abc}
       \bigl[ 
	(\bar{u}t^a \sigma_{\mu\nu}\gamma^5 u)
        (\bar{d}t^b \smn d)(\bar{s}t^cs) +
	(\bar{u}t^a \smn u)(\bar{d}t^b\smn\gamma^5d)(\bar{s}t^cs) \nonumber \\
 &  &  \hspace{0.5cm}
       +(\bar{u}t^a \smn u)(\bar{d}t^b\smn d)(\bar{s}t^c\gamma^5s) +
	(\bar{u}t^a\smn\gamma^5 u)(\bar{d}t^b \smn\gamma^5d)
        (\bar{s}t^c\gamma^5s)  
       \bigr]\nonumber \\
 &   & \hspace{0.5cm}+ (2\,{\it  cyclic\; permutations}\; 
            u\leftrightarrow d \leftrightarrow s)
      {\biggr]}\nonumber \\
 &  &-\frac{9}{40}d^{abc}
       \biggl[  
	(\bar{u}t^a\gamma^5u)(\bar{d}t^bd)(\bar{s}t^cs) +
	(\bar{u}t^a u)(\bar{d}t^b\gamma^5d)(\bar{s}t^cs) +
	(\bar{u}t^a u)(\bar{d}t^bd)(\bar{s}\gamma^5t^cs)   \nonumber \\
 &  &  \hspace{0.5cm}
    +(\bar{u}t^a\gamma^5u)(\bar{d}t^b\gamma^5d)(\bar{s}t^c\gamma^5s) 
        \biggr] \nonumber \\
 &  &- \frac{9}{32}if^{abc}
        \biggl[   
	(\bar{u}t^a \sigma_{\mu\nu}\gamma^5 u)(\bar{d}t^b \sng d)
	(\bar{s}t^c\sgm s) 
       +(\bar{u}t^a \smn u)(\bar{d}t^b\sng\gamma^5d) 
	(\bar{s}t^c\sgm s) \nonumber \\
 &  &  \hspace{0.5cm}
       +(\bar{u}t^a\smn u)(\bar{d}t^b\sng d) 
	(\bar{s}t^c\sgm\gamma^5s)  
       +(\bar{u}t^a\smn\gamma^5 u)
	(\bar{d}t^b\sng\gamma^5d)(\bar{s}t^c\sgm\gamma^5s)  
      \biggr]
   \biggr\}  
\eea
The strategy is the same as in the glueball case. We Fierz-rearrange 
the lagrangian (\ref{eta_c_lagr}) and apply the vacuum dominance 
and PCAC approximations. The coupling of the $\eta_c$ bound state
to the instanton involves the matrix element
\be
\label{l_etac_def}
\lambda_{\eta_c}=\langle\eta_c|g^2G\tilde{G}|0\rangle .
\ee
We can get an estimate of this matrix element using a simple
two-state mixing scheme for the $\eta_c$ and pseudoscalar glueball. 
We write
\bea
\label{mix1}
 |\eta_c\rangle\;\; &=& \;\;\cos(\theta)|\bar{c}c\rangle
                   +\sin(\theta)|gg\rangle, \\
\label{mix2}
 |0^{-+}\rangle &=& -\sin(\theta)|\bar{c}c\rangle
                   +\cos(\theta)|gg\rangle .
\eea
The matrix element $f_{\eta_c}=\langle 0|2m_c\bar{c}\gamma_5 c|
\eta_c\rangle\simeq 2.8\,{\rm GeV}^3$ is related to the charmonium 
wave function at the origin. The coupling of the topological charge
density to the pseudoscalar glueball was estimated using QCD spectral
sum rules, $\lambda_{0^{-+}} = \langle 0|g^2G\tilde{G}|0^{-+}\rangle 
\simeq 22.5\,{\rm GeV}^3$ \cite{Narison:1996fm}. Using the two-state 
mixing scheme the two ``off-diagonal'' matrix elements $f_{0^{-+}}=
\langle 0| 2m_c\bar{c}\gamma_5 c |0^{-+}\rangle$ and $\lambda_{\eta_c}
=\langle 0|g^2G\tilde{G}|\eta_c\rangle$ are given in terms of one 
mixing angle $\theta$. We can estimate this mixing angle by computing 
the charm content of the pseudoscalar glueball using the heavy quark 
expansion. Using \cite{Franz:2000ee}
\be
\label{hqexp}
\bar{c}\gamma^5c = \frac{i}{8\pi m_c}\a_s G\tilde{G} 
    + O\left(\frac{1}{m_c^3}\right),
\ee
we get $f_{0^{-+}}\simeq 0.14\,{\rm GeV}^3$ and a mixing angle 
$\theta \simeq 3^0$. This mixing angle corresponds to 
\be
\label{l_etac}
\lambda_{\eta_c}\simeq 1.12\,{\rm GeV}^3.
\ee 
The uncertainty in this estimate is hard to assess. Below we 
will discuss a perturbative estimate of the instanton coupling 
to $\eta_c$. In order to check the phenomenological consistency 
of the estimate equ.~(\ref{l_etac}) we have computed the 
$\eta_c$ contribution to the $\langle g^2G\tilde{G}(0)g^2G\tilde{G}
(x)\rangle$ correlation function. The results are shown in 
Fig.~\ref{fig_gg}. The contribution of the pseudoscalar
glueball is determined by the coupling constant $\lambda_{0^{-+}}$
introduced above. The couplings of the $\eta$, $\eta'$ and 
$\eta(1440)$ resonances can be extracted from the decays
$J/\psi\to \gamma\eta$ \cite{Novikov:uy}. We observe that 
the $\eta_c$ contribution is strongly suppressed, as one 
would expect. We also show the $\eta_c$ and $0^{-+}$ glueball
contributions to the $\langle \bar{c}\gamma_5 c(0)\bar{c}
\gamma_5 c(x)\rangle$ correlation function. We observe that 
even with the small mixing matrix elements obtained from 
equs.~(\ref{mix1}-\ref{hqexp}) the glueball contribution
starts to dominate the $\eta_c$ correlator for $x>1$ fm. 

  We now proceed to the calculation of the exclusive decay 
rates. There are four final states that contribute to the 
$K\bar{K}\pi$ channel, $\eta_c\to K^+K^-\pi^0$, $K^0\bar{K}^0
\pi^0$, $K^+\bar{K}^0\pi^-$ and $K^-K^0\pi^+$. Using isospin
symmetry it is sufficient to calculate only one of the 
amplitudes. Fierz rearranging equ.~(\ref{eta_c_lagr}) we 
get the interaction responsible for the $\eta_c\to K^+K^-\pi^0$
\be
{\cal L}_{I+A}^{K^+K^-\pi^0} = \int dz \int\frac{d\r}{\r^5}
   d_0(\r)\frac{1}{N_c^2-1}
   \left(\frac{\pi^3\r^4}{\alpha_s^2}\right)
   \left(\frac{4}{3}\pi^2\r^3\right)^3\frac{1}{4}
   (\alpha_s G\tilde{G})
   (\bar{s}\gamma^5u)(\bar{u}\g^5s)(\bar{d}\g^5d).
\ee
The decay rate is given by
\be
\label{KKpi}
 \Gamma_{K^+K^-\pi^0} = 
   \int({\rm phase\;\; space})\x |M|^2 =
\left[ \frac{1}{16\pi\sqrt{2}}A\lambda_{\eta_c}K_\pi K_K^2 \right]^2
 \x \left( 0.111\,{\rm MeV} \right),
\ee
with $A$ given in equ.~(\ref{A_int}). Isospin symmetry implies that 
the other $K\bar{K}\pi$ decay rates are given by
\be
 \Gamma_{K^+K^-\pi^0}
 =\Gamma_{K^0\bar{K}^0\pi^0}
 = \left(\frac{1}{\sqrt{2}}\right)^2 \Gamma_{K^0K^-\pi^+}
 = \left(\frac{1}{\sqrt{2}}\right)^2 \Gamma_{K^+\bar{K}^0\pi^-} .
\ee
The total $K\bar{K}\pi$ decay rate is
\be
 \Gamma_{K\bar{K}\pi}=6\x 
 \left[ \frac{1}{16\pi\sqrt{2}}A\lambda_{\eta_c} K_\pi K_K^2 \right]^2
 \x \left(0.111\,{\rm MeV}\right).
\ee  
In a similar fashion we obtain
\bea
 \Gamma_{\eta\pi\pi}
  &=& \frac{3}{2}\x
   \left[\frac{1}{16\pi}
     A\lambda_{\eta_c} K_\eta^{s}K_\pi^2 \right]^2
     \x \left(0.135\,{\rm MeV}\right),  \\
\Gamma_{\eta'\pi\pi}
  &=& \frac{3}{2}\x
   \left[\frac{1}{16\pi}
   A\lambda_{\eta_c} K_{\eta'}^{s}K_\pi^2 \right]^2
   \x \left( 0.0893\,{\rm MeV}\right),\\
\Gamma_{K\bar{K}\eta}
  &=& 2 \x
   \left[\frac{1}{16\pi}
   A\lambda_{\eta_c} K_{\eta}^{q}K_K^2 \right]^2
   \x \left( 0.0788\,{\rm MeV}\right),\\
\Gamma_{K\bar{K}\eta'}
  &=& 2 \x
   \left[\frac{1}{16\pi}
   A\lambda_{\eta_c} K_{\eta'}^{q}K_K^2 \right]^2
   \x \left( 0.0423\,{\rm MeV}\right),\\
\label{etaetaeta}
\Gamma_{\eta\eta\eta}
  &=& \frac{1}{6} \x
   \left[\frac{3!}{16\pi}
   A\lambda_{\eta_c} (K_{\eta}^{q})^2K_{\eta}^s \right]^2
   \x \left( 0.0698\,{\rm MeV}\right).
\eea
Here, the first factor is the product of the isospin 
and final state symmetrization factors. The second factor 
is the amplitude and the third factor is the phase-space 
integral.

\begin{figure}
\begin{center}
\includegraphics[width=10cm,angle=-90]{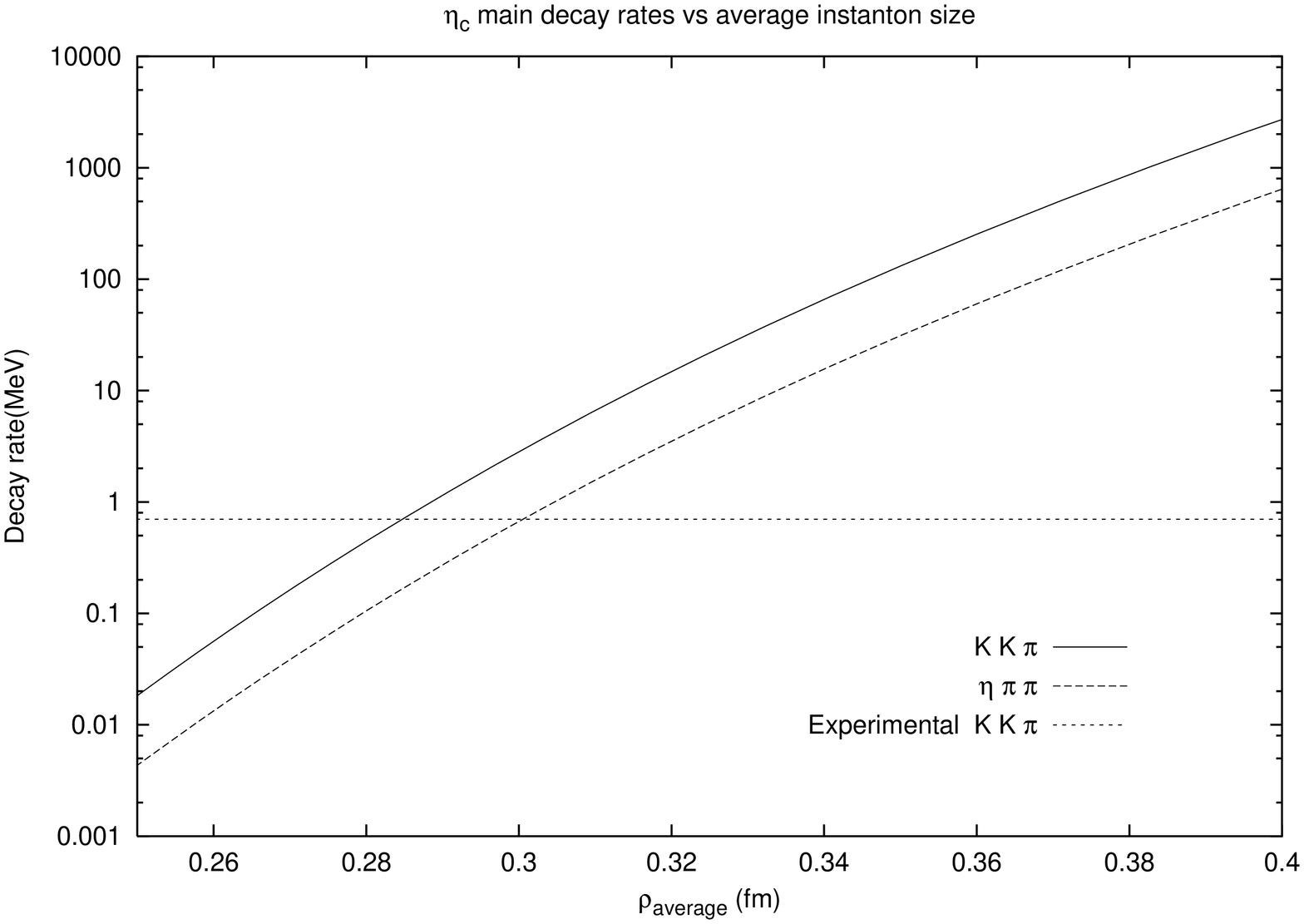}
\end{center}
\begin{center}
\caption{\label{fig_etac_rho}
Decay widths $\eta_c\to KK\pi$ and $\eta_c\to \eta\pi\pi$
as a function of the average instanton size $\rho$. The 
short dashed line shows the experimental $KK\pi$ width.}
\end{center}
\end{figure}

  In Fig.~\ref{fig_etac_rho} we show the dependence of the 
decay rates on the average instanton size. We observe that 
the experimental $K\bar{K}\pi$ rate is reproduced for $\bar{\rho}
=0.29$ fm. This number is consistent with the phenomenological 
instanton size. However, given the strong dependence on the 
average instanton size it is clear that we cannot reliably
predict the decay rate. On the other hand, the following 
ratios are independent of the average instanton size
\bea
 \frac{\Gamma_{K\bar{K}\pi}}{\Gamma_{\eta\pi\pi}} &=&
    4\x \left[
  \frac{K_K^2}
       {\sqrt{2}K_\eta^{s}K_\pi}\right]^2
   \x\left(\frac{0.111}{0.135}\right)=4.23 \pm 1.27 ,   
       \\
 \frac{\Gamma_{\eta\pi\pi}}{\Gamma_{\eta'\pi\pi}} &=&
  \left(\frac{K_\eta^{s}}{K_{\eta'}^{s}} \right)^2
  \x \left(\frac{0.135}{0.0893} \right)=1.01 ,
       \\
 \frac{\Gamma_{K\bar{K}\eta}}{\Gamma_{K\bar{K}\pi}} &=&
  \frac{1}{3}\x \left[
  \frac{\sqrt{2}K_\eta^{q}}
       {K_\pi}\right]^2
   \x\left(\frac{0.0788}{0.111}\right)=0.141 \pm 0.042 , 
       \\
 \frac{\Gamma_{K\bar{K}\eta}}{\Gamma_{K\bar{K}\eta'}} &=&
  \left(\frac{K_\eta^{q}}{K_{\eta'}^{q}} \right)^2
  \x \left(\frac{0.0788}{0.0423} \right)=2.91 ,
       \\ 
 \frac{\Gamma_{\eta\eta\eta}}{\Gamma_{K\bar{K}\pi}} &=&
  \frac{1}{36}\x \left[
  \frac{3!\sqrt{2}(K_\eta^{q})^2K_{\eta}^s}
       {K_\pi K_K^2}\right]^2
   \x\left(\frac{0.0698}{0.111}\right)=0.011 \pm 0.003,
\eea
where we have only quoted the error due to the uncertainty 
in $m_s$. These numbers should be compared to the experimental 
results 
\bea
 \left.\frac{\Gamma_{K\bar{K}\pi}}{\Gamma_{\eta\pi\pi}}
 \right|_{exp} &=&  1.1\pm 0.5\\
\left.\frac{\Gamma_{\eta\pi\pi}}{\Gamma_{\eta'\pi\pi}} 
 \right|_{exp} &=&  1.2\pm 0.6.
\eea
We note that the ratio $B(\eta\pi\pi)/B(\eta'\pi\pi)$
is compatible with our results while the ratio $B(K\bar{K}
\pi)/B(\eta\pi\pi)$ is not. This implies that either there 
are contributions other than instantons, or that the PCAC 
estimate of the ratio of coupling constants is not reliable,
or that the experimental result is not reliable. The branching 
ratios for $\eta\pi\pi$ and $\eta'\pi\pi$ come from MARK
II/III experiments  
\cite{Partridge:1980vk,Baltrusaitis:1985mr}. We 
observe that our results for $B(K\bar{K}\eta)/B(K\bar{K}\pi)$ 
and $B(K\bar{K}\eta')/B(K\bar{K}\pi)$ are consistent with 
the experimental bounds. 

 Another possibility is that there is a significant 
contribution from a scalar resonance that decays into 
$\pi\pi$. Indeed, instantons couple strongly to the 
$\sigma(600)$ resonance, and this state is not resolved
in the experiments. We have therefore studied the direct
instanton contribution to the decay $\eta_c \to \sigma\eta$. 
After Fierz rearrangement we get the effective vertex
\bea
 {\cal L}_{\sigma\eta} &=& \int dA\;
 (\a_s G\tilde{G})\; \frac{1}{4}
 \left[
  (\bar{u}\gamma^5 u)(\bar{d}d)(\bar{s}{s}) +
  (\bar{u} u)(\bar{d}\gamma^5d)(\bar{s}{s}) +
  (\bar{u} u)(\bar{d}d)(\bar{s}\gamma^5{s}) 
 \right] \nonumber \\
&  & \mbox{}-\int dB\;
 m_s (\a_s G\tilde{G})\;\frac{1}{2}
 \left[
  (\bar{u}\gamma^5 u)(\bar{d}d) +
  (\bar{u} u)(\bar{d}\gamma^5d) 
 \right ] ,
\eea
where the integrals $A$ and $B$ are defined in
equ.~(\ref{A_int},\ref{B_int}). The only new matrix element we 
need is $f_\sigma=\langle\sigma|\bar{u}u + \bar{d}d|0\rangle 
\simeq (500\,{\rm MeV})^2$ \cite{Shuryak:1992ke}. We get
\bea
 \Gamma_{\eta_c\to\sigma\eta} &=& \frac{1}{16\pi m_{\eta_c}^3}
  \sqrt{
   [m_{\eta_c}^2 - (m_\sigma + m_\eta)^2]
   [m_{\eta_c}^2 - (m_\sigma - m_\eta)^2]
   } \nonumber \\
  & &\x
  \left[
    \frac{1}{16\pi}f_\sigma \lambda_{\eta_c} 
    [(AQ_s - 2Bm_s)K_\eta^q + AK_\eta^sQ_d]
  \right]^2 .
\eea
Compared to the direct decay $\eta_c\to\eta\pi\pi$ the 
$\eta_c\to\eta\sigma$ channel is suppressed by a factor 
$\sim (2\pi^2/m_{\eta_c}^2)\cdot (Q_qf_\sigma/K_\pi^2)^2
\sim 1/100$. Here, the first factor is due to the difference 
between two and three-body phase space and the second factor
is the ratio of matrix elements. We conclude that the direct 
production of a $\sigma$ resonance from the instanton does
not give a significant contribution to $\eta_c\to\eta(\eta')
\pi\pi$. This leaves the possibility that the $\pi\pi$ 
channel is enhanced by final state interactions.

\begin{figure}
\begin{center}
\includegraphics[width=10cm,angle=0]{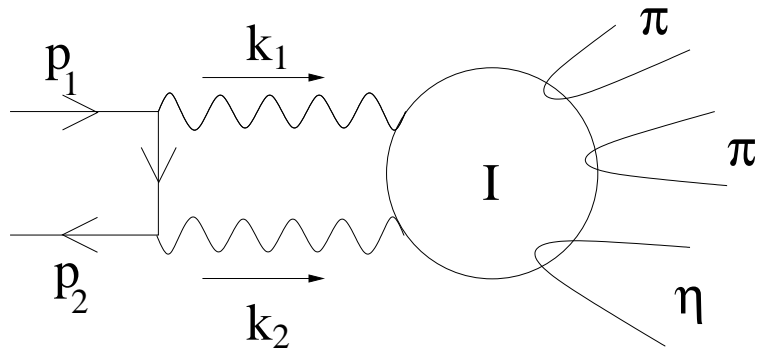}
\end{center}
\begin{center}
\caption{\label{Feynman_diag_fig}
The Feynman diagram corresponding to the perturbative treatment of
charmonium decay.}
\end{center}
\end{figure}

  Finally, we present a perturbative estimate of the coupling
of the $\eta_c$ to the instanton. We follow the method used
by Anselmino and Forte in order to estimate the instanton
contribution to $\eta_c\to p\bar{p}$ \cite{Anselmino:1993bd}.
The idea is that the charmonium state annihilates into two 
gluons which are absorbed by the instanton. The Feynman diagram 
for the process is shown in Fig.\ref{Feynman_diag_fig}. The 
amplitude is given by
\bea
\label{A_ccI}
 A_{c\bar{c}\to I}  &=& 
 g^2 \int \frac{d^4k_1}{(2\pi)^4} \int \frac{d^4k_2}{(2\pi)^4}
 \, (2\pi)^4 \delta^4(p_1+p_2-k_1-k_2) \nonumber \\
 & & \hspace{0.5cm}
   \bar{v}(p_2)\left[ \gamma_\mu \frac{\lambda^a}{2}
   \frac{1}{ {p\!\!\!/}_1-{k\!\!\!/}_1-m_c}
   \gamma_\nu\frac{\lambda^b}{2}\right] u(p_1)\;
   A_\mu^{a,cl}(k_2)A_\nu^{b,cl}(k_1),
\eea
where $u(p)$ and $\bar{v}(p)$ are free particle charm 
quark spinors and $A_\mu^{a,cl}(k)$ is the Fourier transform
of the instanton gauge potential 
\be
\label{A(k)}
 A_\mu^{a,cl}(k) = -i \frac{4\pi^2}{g} 
  \frac{\bar{\eta}^a_{\mu\nu}k^\nu}{k^4}\Phi(k),
  \hspace{1cm} 
  \Phi(k) = 4\left( 1 -\frac{1}{2}K_2(k\rho)(k\rho)^2
    \right).
\ee
The amplitude for the charmonium state to couple to an 
instanton is obtained by folding equ.~(\ref{A_ccI}) with 
the $\eta_c$ wave function $\psi(p)$. In the non-relativistic
limit the amplitude only depends on the wave function at 
the origin. 

  The perturbative estimate of the transition rate is 
easily incorporated into the results obtained above by
replacing the product $A\lambda_{\eta_c}$ in 
equs.~(\ref{KKpi}-\ref{etaetaeta}) according to
\be 
\label{sub}
 A\lambda_{\eta_c} \to 
    \int\frac{d\r}{\r^5}d_0(\r)
    \left(\frac{4}{3}\pi^2\r^3\right)^3
     \left(4\pi\right) \frac{8m_c^{3/2}}{\sqrt{6}}
      |\psi(0)| I_{\eta_c}(\rho)
      \x \frac{g^2(m_c^{-1})}{g^2(\rho)},
\ee 
with 
\be 
\label{I_etac}
 I_{\eta_c}(\rho) = \int d^4k \;
   \frac{\vec{k}^2\Phi(k)\Phi(k-2p_c)}
        {k^4(k-2p_c)^4((k-p_c)^2+m_c^2)}.
\ee
Here, $p_c=(m_c,0)\simeq (M_{\eta_c}/2,0)$ is the momentum 
of the charm quark in the charmonium rest frame. We note that
because of the non-perturbative nature of the instanton 
field higher order corrections to equ.~(\ref{sub}) are 
only suppressed by $g^2(m_c^{-1})/g^2(\rho)$. 

\begin{figure}
\begin{center}
\includegraphics[width=10cm,angle=-90]{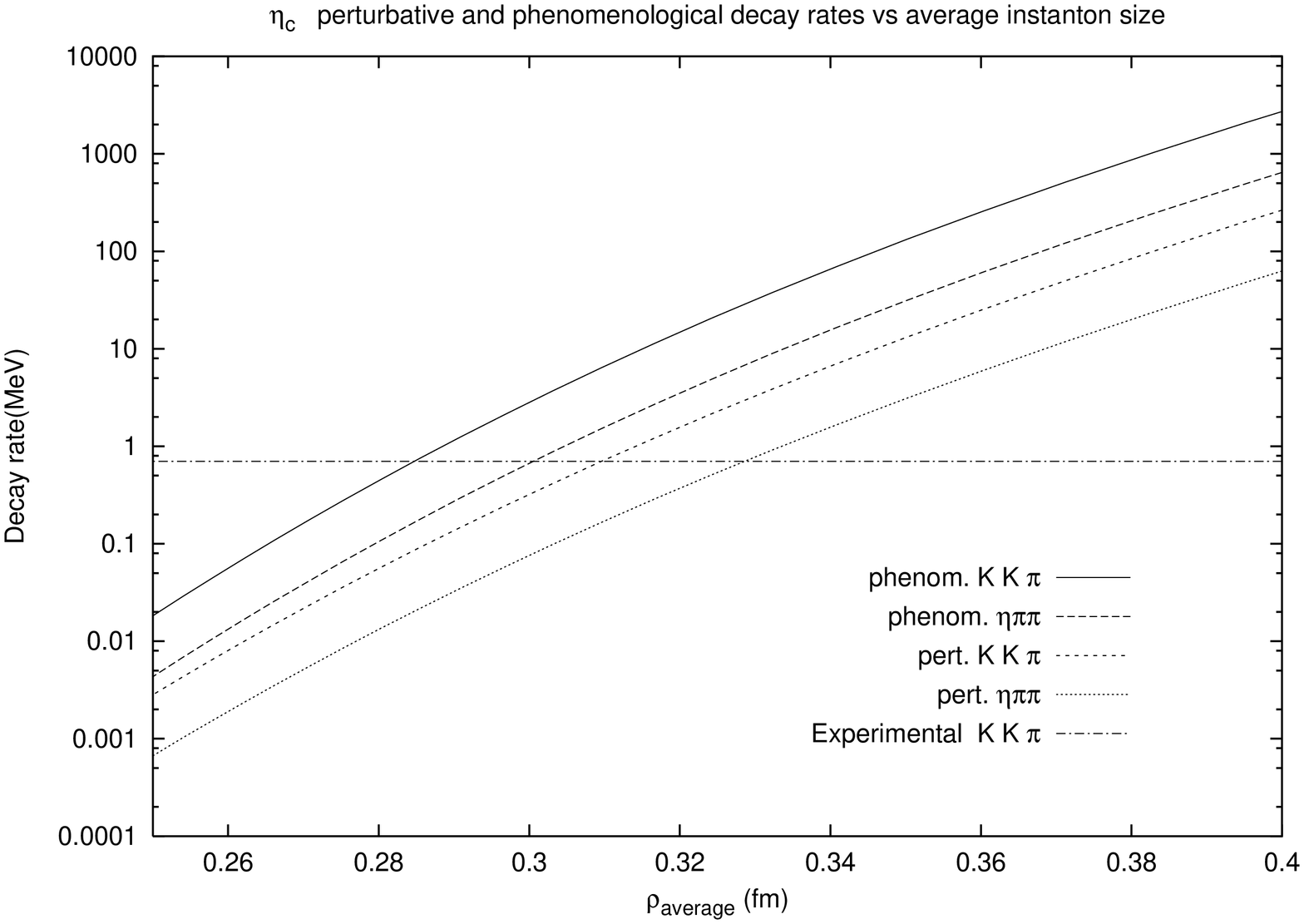}
\end{center}
\begin{center}
\caption{\label{fig_etac_pert}
Decay rates $\Gamma(\eta_c\to K\bar{K}\pi)$ and 
$\Gamma(\eta_c\to \eta\pi\pi)$ as a function of the 
average instanton size $\bar{\rho}$. We show both the 
results from a phenomenological and a perturbative 
estimate of the $\bar{c}c$ coupling to the instanton. }
\end{center}
\end{figure}

The integral $I_{\eta_c}$ cannot be calculated analytically. 
We use the parameterization
\be 
   I_{\eta_c}(\rho) \simeq 
   \frac{\pi^2\;A_0\;\rho^4\log(1+1/(m_c\rho))}
        {1 + B_0\;(m_c\rho)^4\log(1+1/(m_c\rho))},
\ee
which incorporates the correct asymptotic behavior. We
find that $A_0=0.213$ and $B_0=0.124$ provides a good
representation of the integral. In Fig.~\ref{fig_etac_pert} 
we show the results for the $\eta_c$ decay rates as a 
function of the average instanton size. We observe that 
the results are similar to the results obtained from the 
phenomenological estimate equ.~(\ref{l_etac}). The 
effective coupling $(A\lambda_{\eta_c})$ differs from
the estimate equ.~(\ref{l_etac}) by about a factor of 3.
The experimental $K\bar{K}\pi$ rate is reproduced for 
$\bar{\rho}=0.31$ fm.

\section{Chi charm decays}
\label{sec_chi}
 
  Another interesting consistency check on our results is 
provided by the study of instanton induced decays of the 
$\chi_c$ into pairs of Goldstone bosons. The $\chi_c$ is 
a scalar charmonium bound state with mass $m_{\chi_c}=
3415$ MeV and width $\Gamma_{\chi_c}= 14.9$ MeV. In a 
potential model the $\chi_c$ corresponds to the $^3P_0$
state. In perturbation theory the total decay rate is
dominated by $\bar{c}c\to 2g$. The main exclusive decay 
channels are $\chi_c\to 2(\pi^+\pi^-)$ and $\chi_c\to
\pi^+\pi^-K^+K^-$ with branching ratios $(2.4\pm 0.6)\%$
and $(1.8\pm 0.6)\%$, respectively. It would be very 
interesting to know whether these final states are 
dominated by scalar resonances. We will concentrate 
on final states containing two pseudoscalar mesons. 
There are two channels with significant branching ratios,
$\chi_c\to\pi^+\pi^-$ and $\chi_c\to K^+K^-$ with 
branching ratios $(5.0\pm 0.7)\cdot 10^{-3}$ and 
$(5.9\pm 0.9)\cdot 10^{-3}$.

\begin{figure}
\begin{center}
\includegraphics[width=10cm,angle=-90]{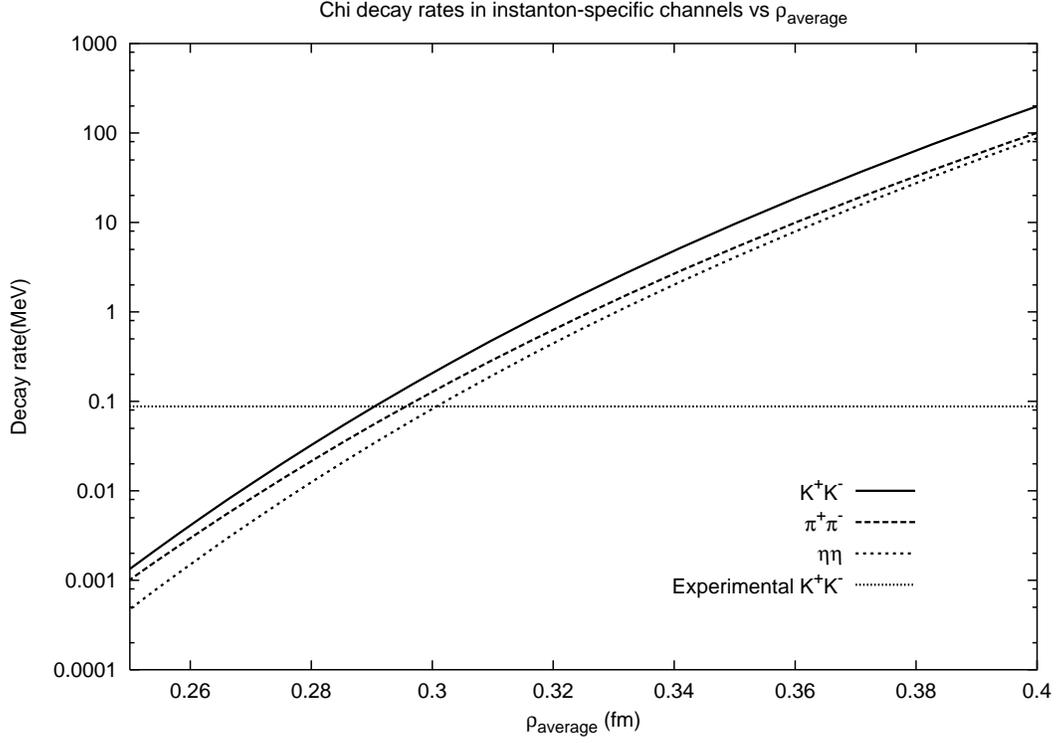}
\end{center}
\begin{center}
\caption{\label{fig_chi_dec}
Decay widths $\chi_c \to K^+K^-,\pi^+\pi^-$ and $\eta\eta$
as a function of the average instanton size $\rho$. The 
short dashed line shows the experimental $K^+K^-$ width.}
\end{center}
\end{figure}

\begin{figure}
\begin{center}
\includegraphics[width=10cm,angle=-90]{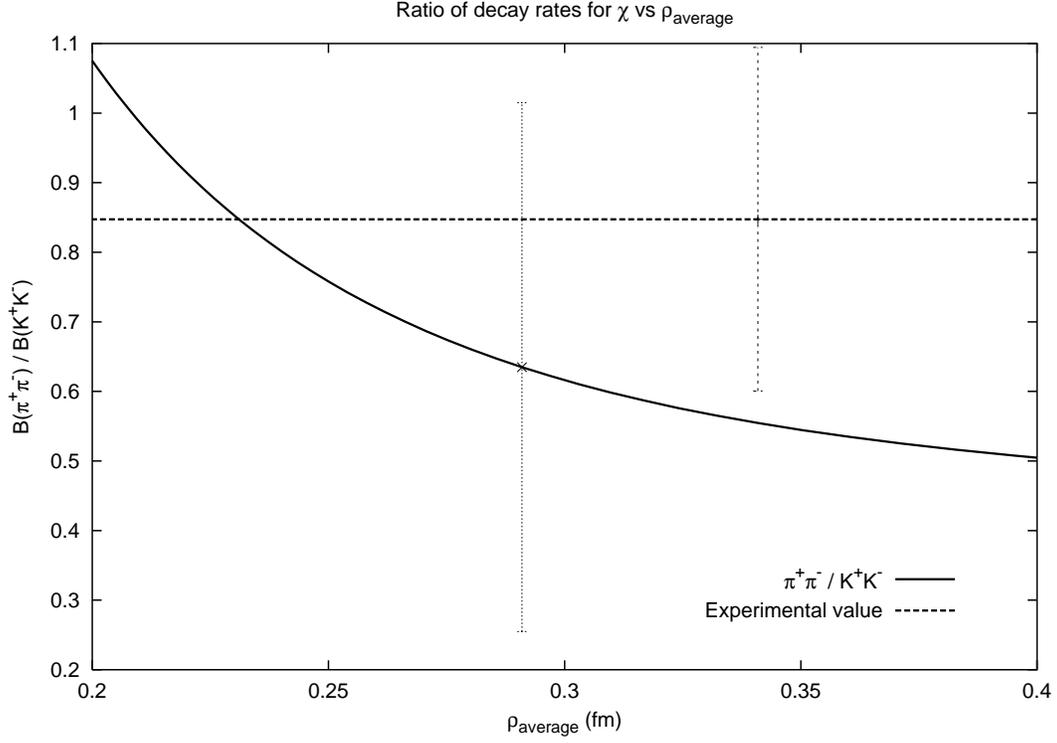}
\end{center}
\begin{center}
\caption{\label{fig_chi_ratio}
Ratio $B(\chi_c\to\pi^+\pi^-)/B(\chi_c \to K^+K^-)$ of 
decay rates as a function of the average instanton size.
The dashed line shows the experimental value 0.84. We
also show the experimental uncertainty, as well as the 
uncertainty in the instanton prediction due to the 
the value of the strange quark mass. }
\end{center}
\end{figure}

 The calculation of these two decay rates proceeds along 
the same lines as the calculation of the $0^{++}$ glueball 
decays. The only new ingredient is the $\chi_c$ coupling to 
the gluon field strength $G^2$. We observe that the total
$\chi_c$ decay rate implies that $\langle 0|2m_c\bar{c}c|\chi_c
\rangle = 3.1\,{\rm GeV}^3\simeq \langle 0|2m_c\bar{c}i\gamma_5c
|\eta_c\rangle$. This suggests that a rough estimate of the
$\chi_c$ coupling to $G^2$ is given by
\be
\lambda_{\chi_c}\equiv \langle \chi_c | g^2 G^2 | 0 \rangle  
 \simeq \lambda_{\eta_c} = 1.12\; {\rm GeV}^3.
\ee
Using this result we can obtain the $\chi_c$ decay rates
by rescaling the scalar glueball decay rates 
equ.~(\ref{glue_pipi_ms}-\ref{glue_etaPetaP_ms}) 
according to 
\be
\Gamma_{\chi_c \rightarrow m1,m2} = 
  \Gamma_{0^{++} \rightarrow m1,m2}  \x 
  \left.   \left( \frac{\lambda_{\chi_c}}{\lambda_{0^{++}}} \right)^2
  \right|_{m_{0^{++}}\to m_{\chi_c}},
\ee
where $m1,m2$ labels the two-meson final state. In Fig.~\ref{fig_chi_dec}
we show the dependence of the $\chi_c$ decay rates on the average 
instanton size $\bar{\rho}$. We observe that the experimental 
$\pi^+\pi^-$ decay rate is reproduced for $\bar{\rho}=0.29$ fm.
In Fig. \ref{fig_chi_ratio} we plot the ratio of decay rates 
for $\pi^+\pi^-$ and $K^+K^-$. Again, the experimental value
is reproduced for $\bar{\rho}\sim 0.3$ fm.

\begin{figure}
\begin{center}
\includegraphics[width=10cm,angle=-90]{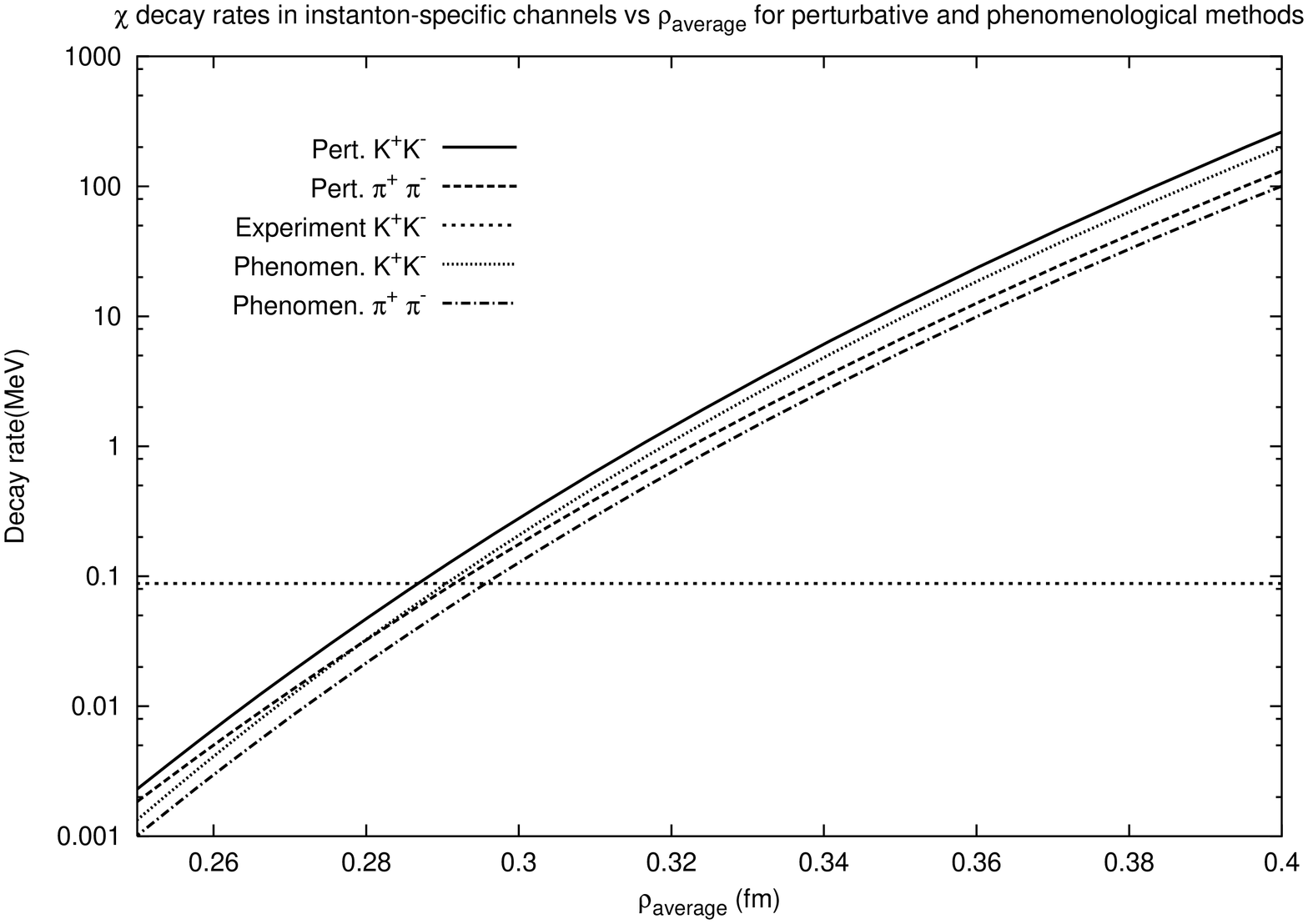}
\end{center}
\begin{center}
\caption{\label{fig_chi_c_pert}
Decay rates $\Gamma(\chi_c\to \pi^+\pi^-)$ and 
$\Gamma(\chi_c\to K^+ K^-)$ as a function of the 
average instanton size $\bar{\rho}$. We show both the 
results from a phenomenological and a perturbative 
estimate of the $\bar{c}c$ coupling to the instanton. }
\end{center}
\end{figure}

  Finally, we can also estimate the $c\bar{c}$ coupling 
to the instanton using the perturbative method introduced
in section \ref{sec_eta}. In the case of the $\chi_c$ we use
\bea
 \frac{1}{4\pi}\lambda_{\chi_c}A 
 &\rightarrow &
  \frac{1}{2\sqrt{3\pi}}
 \sqrt{M_\chi}R'(0)\int \frac{d_0(\rho)}{\rho^5}d\rho
 \left(\frac{4}{3}\pi^2
 \rho^3\right)^3 \frac{g^2(m_c)}{g^2(\rho)}\x I_\chi(\rho),
  \\
 \frac{1}{4\pi}\lambda_{\chi_c}B 
  &\rightarrow & 
 \frac{1}{2\sqrt{3\pi}}
 \sqrt{M_\chi}R'(0)\int \frac{d_0(\rho)}{\rho^5}d\rho
 \left(\frac{4}{3}\pi^2
 \rho^3\right)^2 \rho\, \frac{g^2(m_c)}{g^2(\rho)}\x I_\chi(\rho)\:,
\eea
where $R'(0)\simeq 0.39\,{\rm GeV}^{5/2}$ is the derivative of 
the $^3P_0$ wave function at the origin and $I_{\chi_c}$ is the 
loop integral
\be
  I_\chi(\rho)=\int\! d^4 k\: \frac{\Phi(k)\Phi(|2p_c-k|)}
{k^4(2p_c-k)^4}\:\:
 \frac{15(k-p_c)^2 +3m_c^2 + 4\vec{k}^2 }{(k-p_c)^2+m_c^2}.
\ee
In Fig.~\ref{fig_chi_c_pert} we compare the perturbative 
result with the phenomenological estimate. Again, the results
are comparable. The experimental $\pi^+\pi^-$ rate is reproduced
for $\bar{\rho}=0.29$ fm.

\section{Summary}
\label{sec_sum}
 
 In summary we have studied the instanton contribution to
the decay of a number of ``gluon rich'' states in the 
(1.5-3.5) GeV range, the scalar glueball, the $\eta_c$ and
the $\chi_c$. In the case of charmonium instanton induced
decays are probably a small part of the total decay rate, 
but the final states are very distinctive. In the case 
of the scalar glueball classical fields play an important 
role in determining the structure of the bound state and 
instantons may well dominate the total decay rate. 

 We have assumed that the gluonic system is small and that 
the instanton contribution to the decay can be described
in terms of an effective local interaction. The meson 
coupling to the local operator was determined using PCAC.
Using this method we find that the scalar glueball decay 
is dominated by the $K\bar{K}$ final state for glueball
masses $m_{0^{++}}>1$ GeV. In the physically interesting 
mass range $1.5\,{\rm GeV}<m_{0^{++}}<1.75\,{\rm GeV}$
the branching ratios satisfy $B(\eta\eta):B(\pi\pi):
B(\bar{K}K)=1:(3.3\pm 0.3):(5.5\pm 0.5)$.
 
 Our main focus in this work are $\eta_c$ decays into 
three pseudoscalar Goldstone bosons. We find that the 
experimental decay rate $\Gamma(\eta_c\to K\bar{K}\pi)$
can be reproduced for an average instanton size $\bar{\rho}
=0.31$, consistent with phenomenological determinations
and lattice results. This in itself is quite remarkable, 
since the phenomenolgical determination is based on
properties of the QCD vacuum.

The ratio of decay rates $B(\eta'\pi\pi):B(\eta\pi\pi):B(K\bar{K}\pi)
=1:1:(4.2\pm 1.3)$ is insensitive to the average 
instanton size. While the ratio $B(\eta'\pi\pi):B(\eta
\pi\pi)=1:1$ is consistent with experiment, the ratio
$B(\eta\pi\pi):B(K\bar{K})=1:(4.2\pm 1.3)$ is at best 
marginally consistent with the experimental value
$1.1\pm 0.5$. We have also studied $\chi_c$ decays 
into two pseudoscalars. We find that the absolute decay 
rates can be reproduced for $\bar{\rho}=0.29$ fm. 
Instantons are compatible with the measured ratio
$B(K^+K^-):B(\pi^+\pi^-)=1.2$

  There are many questions that remain to be answered. 
On the experimental side it would be useful if additional
data for the channels $\eta_c\to\eta'\pi\pi,\eta\pi\pi$
were collected. One important question is whether $(\pi\pi)$
resonances are important. It should also be possible to 
identify the smaller decay channels $\eta_c\to K\bar{K}\eta, 
K\bar{K}\eta'$. In addition to that, it is interesting to 
study the distribution of the final state mesons in all
three-meson channels. Instantons predict that the production 
mechanism is completely isotropic and that the final state 
mesons are distributed according to three-body phase space.

  In addition to that, there are a number of important 
theoretical issues that remain to be resolved. In the limit
in which the scalar glueball is light the decay $0^{++}\to \pi\pi
(\bar{K}K)$ can be studied using effective lagrangians based on 
broken scale invariance \cite{Shifman:1988zk,Jaminon:ac,Jin:2002up}.
Our calculation based on direct instanton effects is valid
in the opposite limit. Nevertheless, the instanton liquid 
model respects Ward identities based on broken scale invariance
\cite{Schafer:1996wv} and one should be able to recover the
low energy theorem. In the case $0^{++}\to \pi\pi(\bar{K}K)$ 
one should also be able to study the validity of the PCAC 
approximation in more detail. This could be done, for example,
using numerical simulations of the instanton liquid. Finally 
we need to address the question how to properly compute the 
overlap of the initial $\bar{c}c$ system with the instanton. 
This, of course, is a more general problem that also affects 
calculations of electroweak baryon number violation in high 
energy $p\bar{p}$ collisions \cite{Ringwald:1989ee,Espinosa:qn} 
and QCD multi-particle production in hadronic collisions 
\cite{Nowak:2000de}.

Acknowledgments: We would like to thank  D.~Kharzeev, E.~Shuryak 
and A.~Zhitnitsky for useful discussions. This work was supported 
in part by US DOE grant DE-FG-88ER40388 and by a DOE-OJI grant.


\newpage

\end{document}